\documentclass[useAMS,usenatbib]{mn2e}
\usepackage{times}
\usepackage{amssymb}
\usepackage{times}
\usepackage{txfonts}
\usepackage{bm}
\usepackage{color}
\usepackage{epsfig}

\title[A Disk Wind Model for BALQSOs]{A Simple Disk Wind Model for Broad Absorption Line Quasars}

\author[Higginbottom et al.]{N.~Higginbottom,$^1$\thanks{nick\_higginbottom@fastmail.fm} C.~Knigge,$^1$ K.~S.~Long,$^2$ S.~A.~Sim,$^3$ and J.~H.~Matthews$^1$
\medskip  
\\$^1$School of Physics and Astronomy, University of Southampton, Highfield, Southampton, SO17 1BJ, United Kingdom
\\$^2$Space Telescope Science Institute, 3700 San Martin Drive, Baltimore, MD, 21218
\\$^3$School of Mathematics and Physics, Queens University Belfast, University Road, Belfast, BT7 1NN, Northern Ireland, UK}

\date{\today}

\pagerange{\pageref{firstpage}--\pageref{lastpage}}
\pubyear{2013}

\newcommand\plotone[1]{\centering\includegraphics[width=\hsize]{#1}}

\newcommand{\ltappeq}{\raisebox{-0.6ex}{$\,\stackrel
{\raisebox{-.2ex}{$\textstyle <$}}{\sim}\,$}}
\newcommand{\gtappeq}{\raisebox{-0.6ex}{$\,\stackrel
{\raisebox{-.2ex}{$\textstyle >$}}{\sim}\,$}}

\begin{document}
\maketitle

\label{firstpage}

\begin{abstract}
 
Approximately $20\%$ of quasi-stellar objects (QSOs) exhibit broad,
blue-shifted absorption lines in their ultraviolet spectra. Such
features provide clear evidence for significant outflows from these
systems, most likely in the form of accretion disk winds. These winds
may represent the ``quasar'' mode of feedback that is often invoked in
galaxy formation/evolution models, and they are also key to
unification scenarios for active galactic nuclei (AGN) and QSOs. 
To test these ideas, we construct a simple benchmark model of an
equatorial, biconical accretion disk wind in a QSO and use a Monte Carlo 
ionization/radiative transfer code to calculate 
the ultraviolet spectra as a function of viewing angle. We find that
for plausible outflow parameters, sightlines looking directly
into the wind cone do produce broad, blue-shifted absorption features
in the transitions typically seen in broad absorption line
QSOs. However, our benchmark model is intrinsically X-ray weak in
order to prevent overionization of the outflow, and the wind does 
not yet produce collisionally excited line emission at the level
observed in non-BAL QSOs. As a first step towards addressing these
shortcomings, we discuss the sensitivity of our results to changes in
the assumed X-ray luminosity and mass-loss rate, $\dot{M}_{wind}$. In
the context of our adopted geometry, $\dot{M}_{wind} \sim
\dot{M}_{acc}$ is required in order to produce significant
BAL features. The kinetic luminosity and momentum carried by such
outflows would be sufficient to provide significant feedback.

\end{abstract}

\begin{keywords}
galaxies: active - quasars: general - quasars: absorption lines - radiative transfer - 
methods: numerical 
\end{keywords}

\section{Introduction}
\label{introduction}
Outflows are key to our understanding
of quasi-stellar objects (QSOs) and active galactic nuclei
(AGN). First, they are ubiquitous \citep{ganguly_08, kellerman_89}, 
suggesting that they are an
integral part of the accretion process that fuels the central
supermassive black hole (SMBH). Second, they can 
substantially alter -- and in some cases dominate -- the observational
appearance of QSO/AGN across the entire spectral domain, from X-rays 
\citep[e.g.][]{turner_miller_09}, through the ultraviolet band \citep[e.g.][]{weyman}, 
to radio wavelengths \citep[e.g.][]{begelman_84}. Third, they represent a ``feedback'' 
mechanism, i.e. a means by
which the central engine can interact with its galactic and
extragalactic environment \citep{king_03,king_05,fabian_12}. 

There appear to be two distinct modes of mass loss from
AGN/QSOs: (i) highly collimated relativistic jets 
driven from the immediate vicinity of the SMBH, (ii) slower-moving,
less collimated, but more heavily mass-loaded 
outflows from the surface of the accretion disk surrounding
the SMBH. Intriguingly, both types of mass loss 
are also seen in other types of accreting systems on all astrophysical
scales, such as young stellar 
objects \citep[YSOs; e.g.][]{lada_85}, accreting white dwarfs in cataclysmic 
variables \citep[CVs; e.g.][]{kording_11,cordova_mason_82} and
X-ray binaries containing neutron stars or stellar-mass black holes
\citep[e.g.][]{ponti_fender,fender_06}. Thus the connection between
accretion and mass loss appears to be a fundamental and universal
aspect of accretion physics.

Both jets and winds have been invoked as important feedback
mechanisms. The current consensus \citep[e.g.][]{fabian_12}
appears to be that winds may dominate feedback during bright QSO
phases (the so-called 
``quasar-mode'' of feedback), while the kinetic power of jets may
dominate during less active phases (``radio
mode''). More specifically, the quasar mode may be responsible for 
quenching the burst of star formation that characterizes the initial
growth phase of a massive galaxy, and thus for moving galaxies
from the actively star-forming ``blue cloud'' to the more passively
evolving ``red sequence'' \citep{granato_01,schawinski_07}. 
Conversely, the radio mode appears to be responsible for inhibiting 
new star formation from taking place via cooling flows in massive red
galaxies in the centres of clusters and groups \citep{mcnamara_nulsen}. 

Of the two outflow/feedback modes, jets are arguably the better
understood, at least phenomenologically. Powerful jets tend to
announce their presence via strong radio (and X-ray) emission, and
they can often be imaged directly. By contrast, much less is known about the
properties of disk winds. This is at least in part due to their much
smaller physical size \citep[e.g.][]{murray_chiang_96}, which usually
prevents direct imaging studies\footnote{Some disk winds in the local universe {\em have} been
imaged directly. For example, the spatially resolved ``ruff'' in the
radio emission of the classic microquasar SS433 is thought to arise in
a disk wind \citep{blundell_01}.}. As a result, the geometry, kinematics
and energetics of disk winds typically have to be inferred by indirect
means \citep[e.g.][]{slos_vit, kwd,arav_08,neilsen_09}.

Arguably the most direct evidence of disk winds in
AGN/QSO is provided by the class of broad absorption line quasars
(BALQSOs). These objects make up $\simeq\!20\%$ of the QSO population
(\citealt{knigge_balfrac}; but also see \citealt{allen_2011}) and exhibit
broad, blue shifted absorption lines and/or P-Cygni profiles
associated with strong resonance lines in the ultraviolet (UV)
waveband. Such features are classic signatures of outflows and are
also seen, for example, in hot stars \citep[e.g.][]{morton_67} and CVs
\citep[e.g.][]{greenstein_oke}. 

In the majority ($\gtrsim 90\%$) of BALQSOs, the so-called HiBALs, the observed
absorption features are all due to relatively high-ionization species
such as C~\textsc{iv}, Si~\textsc{iv} and N~\textsc{v}. In a
smaller ($\simeq 5\%$) sub-set of BALQSOs, the so-called LoBALs,
blueshifted absorption is also observed in transitions associated with
lower ionization species, such as Mg~\textsc{ii} and Al~\textsc{ii}.
Finally, a small number ($\sim1\%$) of extreme cases, the FeLoBALS, also
show absorption in Fe~\textsc{ii} and Fe~\textsc{iii}.

Intriguingly, these are the same transitions that are typically
seen in non-BAL QSOs and AGN, except that here they appear as pure
(but still broad) emission lines. Moreover, despite differences in the 
continuum spectral energy distributions (SEDs), it appears that
 both BAL and non-BAL QSOs are drawn from the same parent population
 \citep{reichard}. These considerations have naturally led to the
 suggestion that the apparent 
differences between BALQSOs and non-BAL QSOs are merely orientation
effects \citep{elvis_2000}. In such a geometric unification scenario, the 
``broad emission line region'' (BELR) in non-BAL QSOs is a (possibly different) 
part of the same
accretion disk wind that produces the absorption features in
BALQSOs. The observational differences between the two classes are
then solely due to viewing angle. BAL features are observed when the
UV-bright accretion disk is viewed through the outflow,
while only BELs are seen when the system is viewed from other
directions.

Given the central role played by disk winds in AGN/QSO unification
scenarios and galaxy evolution models, it is clearly
important to determine the physical properties -- i.e. the geometry,
kinematics and energetics -- of these outflows. However, while there
have been many previous efforts in this direction
\citep{murray_95,bottorff_97,everett_01,everett_02,
  proga_kallman,sim_05,schurch_07,schurch_09,borguet_10,giustini_proga}, 
there has not really been a convergence towards a unique physical
description. We have therefore embarked on a project to
construct a comprehensive, semi-empirical picture of disk winds in
AGN/QSO. Our ultimate goal is to test if outflows
can account simultaneously for most of the diverse
observational tracers of disk winds in AGN/QSOs. 
The present paper represents the first step in this long-term
program. Our aim here is to test if a simple, physically motivated disk
wind model can give rise to spectra that resemble those of BALQSOs when
the line of sight towards the central engine lies within the wind
cone. 

\begin{figure*}
\plotone{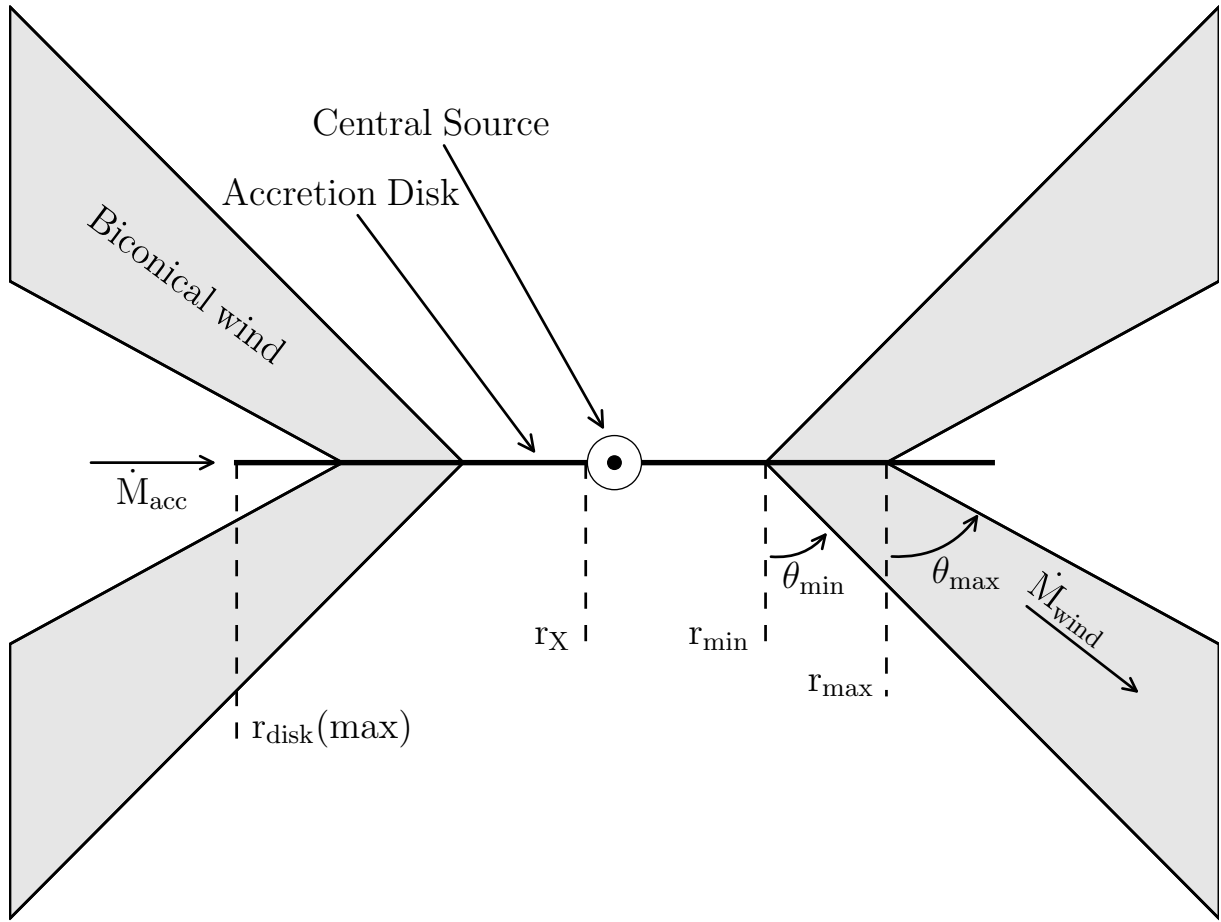}
\caption{A sketch illustrating the main features of our kinematic disk
  wind model.}
\label{sv_cartoon}
\end{figure*}

The detailed plan of this paper is as follows. In Section~\ref{kin_mod}, we
describe the family of kinematic disk wind models we use to describe AGN/QSO
outflows. In Section~\ref{python_details},
we briefly describe our Monte Carlo ionization and radiative transfer code,
\textsc{python} \citep[][LK02 hereafter]{long_knigge}, focusing on the
extensions and modifications we have implemented to enable its
application to AGN/QSO winds. In Section~\ref{fid_mod}, 
we discuss observational
and theoretical constraints that restrict the relevant parameter
space of the model. These considerations allow us to define a benchmark disk wind
model as a reasonable starting point for investigating the
impact of disk winds on the spectra of AGN/QSO. In
Section~\ref{results}, we present and discuss the synthetic spectra
produced by this model. In Section~\ref{discussion}, we consider the
main shortcomings of the benchmark model and, as a first step towards
overcoming them, explore the model's sensitivity to changes in X-ray
luminosity and mass outflow rate. Finally in Section~\ref{conclusion},
we summarise our findings.

\section{The Kinematic Disk Wind Model}
\label{kin_mod}
Since the driving mechanism, geometry and dynamics of AGN/QSO disk
winds are all highly uncertain, we use a flexible, purely kinematic
model to describe the outflow. This allows us to describe a wide range
of plausible disk winds within a single, simple framework. 
The specific prescription we
use was developed by \citet[SV93 hereafter]{slos_vit} 
to model the accretion disk winds observed in CVs, i.e. accreting
white dwarf binary systems.  

The geometry of our outflow model is illustrated in Figure
\ref{sv_cartoon}. A biconical wind is taken to
emanate between radii $r_{min}$ and $r_{max}$ in the accretion
disk, with the wind boundaries making angles of $\theta_{min}$ and
$\theta_{max}$ with the axis of symmetry. At each radius, $r_0$, within this 
range, the wind leaves the disk with a poloidal (non-rotational)
velocity vector oriented at an angle $\theta$ to the axis of
symmetry, with $\theta$ given by 
\begin{equation}
\theta=\theta_{min}+\left(\theta_{max}-\theta_{min}\right)x^{\gamma},
\end{equation}
where
\begin{equation}
x=\frac{r_0-r_{min}}{r_{max}-r_{min}}.
\end{equation}
The parameter $\gamma$ can be used to vary the concentration of
poloidal streamlines towards either the inner or outer regions of the wind, but
throughout this work we fix it to 1, so the poloidal streamlines are
equally spaced in radius. 

The poloidal velocity, $v_l$, along a streamline in our model is given
by 
\begin{equation}
v_l=v_0+\left[v_{\infty}(r_0)-v_0\right]\frac{\left(l/R_v\right)^{\alpha}}{\left(l/R_v\right)^{\alpha}+1},
\label{v_law}
\end{equation}
where $l$ is distance measured along a poloidal streamline. This
power-law velocity profile was adopted by SV93 in order to give a 
continuous variation in the derivative of the velocity and a 
realistic spread of Doppler-shifted frequencies in the outer portion
of the wind (${l > R_v})$. We have similar requirements.
The initial poloidal velocity of the wind, $v_0$ is (somewhat
arbitrarily) set to $6~\rm{km~s^{-1}}$ for all streamlines, comparable
to the sound speed in the disk photosphere. The
wind then speeds up on a characteristic scale length $R_v$, 
defined as the position along the poloidal streamline at which the
wind reaches half its terminal velocity, $v_{\infty}$.
The terminal poloidal velocity along each streamline is set to a fixed
multiple of the escape velocity at the streamline foot point, 
\begin{equation}
v_{\infty}=fv_{esc},
\label{v_infty}
\end{equation}
so the
innermost streamlines reach the highest velocities.
The power-law index $\alpha$ controls the shape of the velocity
law: as $\alpha$ increases, the acceleration is increasingly
concentrated around $l = R_v$ along each streamline. 
For large $\alpha$, the initial poloidal velocity
stays low near the disk and then increases quickly through $R_v$ to 
values near $v_\infty$. 

The wind is assumed to initially share the Keplerian rotation profile
of the accretion disk, i.e. $v_{\phi,0}(r_0) = v_{K}(r_0)$ at the base
of the outflow.  As the wind rises above the disk and expands, we
assume that specific angular momentum is conserved,
so that the rotational
velocities decline linearly with increasing cylindrical distance from
the launch point.

\begin{figure}
\plotone{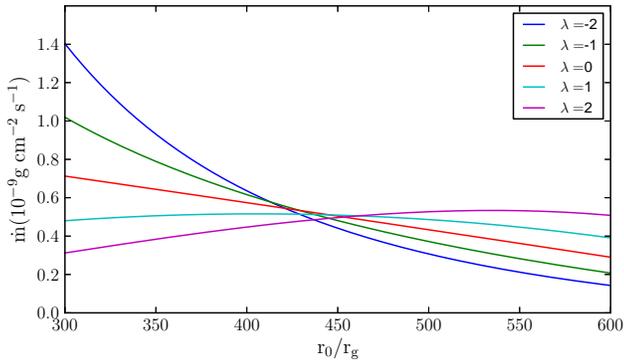}
\caption{Dependence of mass loss rate per unit area on the disk 
on radius for three values of
$\lambda$. $\dot{M}$ and $\theta(r_0)$ are set as in the benchmark model 
defined in Section~\ref{fid_mod} }
\label{lambda_plot}
\end{figure}

The SV93 prescription also allows us to control the mass
loading of the outflow as a function of radius. More specifically, the
local mass-loss rate per unit surface area on the disk ($\dot{m}_{wind}$) 
is given by 
\begin{equation}
\dot{m}_{wind}\propto\dot{M}_{wind}r_0^{\lambda}\cos\left[\theta(r_0)\right]
  \label{lambda_eqn}
\end{equation}
where $\dot{M}_{wind}$ is the global mass loss rate through the wind.
The run of mass loss rate per unit area with $r_0$ is illustrated in Figure~\ref{lambda_plot}.

The combination of local mass-loss rate, launching angle and poloidal
velocity is sufficient to uniquely determine the density at any point
in the wind via a continuity equation (see SV93 for details). The
resulting wind is a smooth, single phase outflow without clumps. 
There is evidence for structure in BALQSO outflows,
apparent from complex line shapes  \citep[e.g.][]{ganguly_06,simon_10}
and time variability \citep[e.g.][]{capellupo_11,capellupo_12,capellupo_13}, but
this is something our simple model cannot address. However, hot star
winds and CV winds also exhibit variability and small-scale structure,
yet much has been learned about these winds by focussing on the
global, underlying smooth flow field. This is the approach followed
here. 

\section{Ionization, Radiative Transfer and Spectral Synthesis}
\label{python_details}

In order to calculate the spectra predicted to emerge from our disk
winds, we use \textsc{python}, a radiative
transfer code designed to model biconical outflows and  first
described by LK02. The code has been used previously to calculate
model spectra for disk-dominated CVs \citep{noebauer_10} and young
stellar objects \citep{sim_drew_05}. Here, we provide a brief
overview of the code, focussing particularly on the modifications we
have made recently to enable modelling of BALQSOs. All simulations
shown in this paper were carried out with version 75 of
\textsc{python}.

\subsection{Basic Structure}

\textsc{python} is a
hybrid Monte Carlo/Sobolev radiative transfer code that works by
following the progress of energy packets through a simulation grid of
arbitrary size, shape and discretization. For the simulations described 
here, we utilize an azimuthally symmetric
cylindrical grid.  Energy packets are
characterized by a frequency and a weight, defined so that
the sum of all packets correctly represents the luminosity and
spectral energy distribution of all radiation sources in the model. 

The thermal and ionization structure of a wind model is computed 
iteratively through a series of `ionization cycles'. During each cycle, 
energy packets are tracked through the wind and their effect on 
the wind's temperature and ionization state is computed (see Section
\ref{ioniz_calc} below). Once the ionization structure of the wind has been determined,
a series of `spectral cycles' are executed, in which synthetic spectra are generated at specific 
inclination angles.  
 
\subsection{Radiation Sources}
\label{rad_sources}

For simulations of AGN/QSO, the radiative sources included are a
central X-ray source, an accretion disk and the wind itself. We assume
that the central source produces a power-law SED and that the optically
thick, geometrically thin disk radiates as an ensemble of blackbodies. 
We do not at present apply the colour temperature
correction as suggested in \cite{done_12} but the correction
is in any case modest for the accretion disk in our benchmark model.
 
The geometry of the X-ray emitting region in AGN is not fully
understood, but the emission is believed to arise from a hot
corona above the inner regions of the accretion disk
\citep[e.g.][]{galeev}. Here, we follow 
\cite{sim_2} and assume the central X-ray source 
is an isotropically
radiating sphere, with radius equal to the 
innermost radius of the accretion disk (typically set to $r_{ISCO} =
6r_g = 6 G M_{BH} / c^2$, the radius of the
innermost circular stable orbit [ISCO] for a non-rotating black hole). The
X-ray source  produces energy
packets governed by a power law in flux, i.e. 
\begin{equation}
F_X({\nu})=K \nu^{\alpha_x}, 
\end{equation}
where $\alpha_x$ is the spectral index. In practice, we characterize
the X-ray source primarily by its integrated luminosity between
2-10~keV, which is related to $K$ and $\alpha$ via 
\begin{equation}
L_{X}=\int_{\rm{2~keV}}^{\rm{10~keV}}{K\nu^{\alpha_x}d\nu}.
\end{equation}

UV continuum emission in AGN is thought to be dominated by an
accretion disk.  We assume the disk has the radial temperature profile
of the standard thin disk  
equations \citep{shakura_sunyaev}, i.e. 
\begin{equation}
T_{d,eff}(r_0) = T_\star
\left(\frac{r_{BH}}{r_0}\right)^{3/4}\left(1-\sqrt{\frac{r_{BH}}{r_0}}\right)^{1/4},
\end{equation}
where 
\begin{equation}
T_{\star} = \left(\frac{3 G M_{BH} \dot{M}_{acc}}{8\sigma\pi r_{BH}^3}\right)^{1/4}.
\end{equation}
Here, $\dot{M}_{acc}$ is the accretion rate through the disk, and we
take $r_{BH} = r_{ISCO}$, appropriate for a
non-rotating SMBH.  

Finally, the wind itself is a hot thermal plasma and therefore
radiates through various atomic processes including line emission,
recombination and/or free-free processes. Unlike the other radiation
sources, the radiation field produced by the wind has to be computed
iteratively, since it depends on the ionization state and temperature
of the outflow. The wind is always assumed to be in radiative
equilibrium, i.e. it reprocesses radiation originally produced
by the disk and the central source. These components therefore set the
total luminosity of the system -- the wind is not a \emph{net} source
of photons in our model.
\footnote{In principle, it would be simple to allow for
non-radiative energy input into the wind (e.g. shocks or energy
release via magnetic reconnection events). However, the relevance of 
these processes, let alone the detailed form they would take, is
presently unknown.}

\subsection{Thermal and Ionization Equilibrium}
\label{ioniz_calc}

\subsubsection{Overall Iteration Scheme}
\label{itscheme}

At the beginning of each simulation, the electron temperature ($T_e$) and
ionization state of the wind are
initialized by assuming a reasonable value for $T_e$ and
adopting LTE ionization fractions. As the energy packets
progress through the wind, they lose energy via continuum and line
absorption, thereby heating the wind. They can also change frequency 
in the observer frame via scattering in the outflowing wind. However, since
such scattering events are assumed to be coherent in the comoving frame, 
these frequency changes do not contribute to heating.

As energy packets move through the wind, all of the quantities that are needed 
to calculate the ionization are logged. These include 
the frequency-integrated mean (angle-averaged) 
intensity ($J$), the mean frequency ($\overline{\nu}$) and the
standard deviation ($\sigma_{\nu}$) of the frequency of energy packets passing
through the cell. These quantities are logged in a series of frequency bands to 
provide information for the ionization calculations described
below. We also calculate `global' values for these quantities 
(integrated over the entire frequency range) in each cell, 
for use in the heating and cooling calculations. All energy packets
are tracked until they escape the wind.

After each cycle in the ionization stage, a new estimate of $T_e$
is computed via a process of minimising the absolute value of the difference 
between $H$, 
the heating rate computed during the cycle, and $C(T_e)$, the cooling
rate which is expressed as a function of $T_e$. The code then recomputes  
the ionization state of the wind, using
the new estimate of the temperature and the information recorded
about the radiation field in the cell. In order to improve
the stability of the code, we restrict the 
amount by which $T_e$ can change in one ionization cycle.

Armed with new estimates of electron temperature and ionization state,
the code then starts the next ionization cycle. This iteration 
proceeds until the electron temperature in each cell has converged to
a stable value. At this point, all physical parameters of the wind
are frozen, and  synthetic spectra can be produced for any desired wavelength range or inclination angle.

\subsubsection{Ionization Balance}

LK02 determined the ionization state of the wind using a ``modified
on-the spot approximation'' first derived  by \cite{abbot_lucy}. This treatment
is designed for a situation in which  the local radiation field is fairly close to
a dilute blackbody, as is the case for O-star winds and CVs. However,
unlike O stars and CVs, AGN/QSO emit a significant fraction of their
energy in X-rays and have global SEDs that are poorly described by 
blackbodies.

We have therefore implemented a new method to calculate the ionization
state of the wind into \textsc{python}, which is better suited to
situations in which relatively soft (UV) and relatively hard (X-ray)
components both contribute significantly to the ionizing radiation field. 
Our new ionization treatment follows that described by
\cite{sim_2}, which can be summarized by the ionization equation 
\begin{equation}
\frac{n_{i+1,0}n_e}{n_{i,0}}= \Phi^*_{T_e} \zeta(T_e)
S_i(T_e,J_{\nu}).
\label{ioniz_2}
\end{equation}
Here, $n_e$ is the electron density and $n_{i,0}$ represents the
density of the ground state of ionization stage $i$ of a particular
atomic species. The quantity $\Phi^*_{T_e}$ is the ratio
${n_{i+1,0}n_e}/{n_{i,0}}$ computed 
via the Saha equation at temperature $T_e$, while $\zeta(T_e)$ is the
fraction of recombinations going directly into the ground state,
allowing for both radiative recombinations into all levels and
dielectronic recombinations. The correction factor $S_i(T_e,J_{\nu})$
is the ratio of the photoionization rate expected for the actual SED
to that which would be produced by a blackbody at $T_e$,
\begin{equation}
S_i(T_e,J_{\nu})= \frac{\displaystyle{\int_{\nu_0}^{\infty}}
  {J_{\nu}\sigma_i(\nu)\nu^{-1}d\nu}
}{\displaystyle{\int_{\nu_0}^{\infty}}{B_{\nu}(T_e)\sigma_i(\nu)}\nu^{-1}d\nu}.
\label{i_calc}
\end{equation}
Here, $\sigma_i(\nu)$ is the frequency-dependent photoionization
cross-section for ionization state $i$, and $J_{\nu}$ is the
monochromatic mean intensity. 

This formulation allows us to correctly account for the
photo-ionizing effect of radiation fields with arbitrary SEDs,
provided there are sufficient photon packets to 
adequately characterise the SED. In this work, we model the SED in
each cell after each ionization cycle by splitting it into a series of
user-defined bands. Equation \ref{i_calc} can then be
re-written as 
\begin{equation}
S_i(T_e,J)= 
\frac{\displaystyle{\sum_{band~j=0}^{n}}~{\int_{\nu_j}^{~\nu_j+1}{J_{\nu,j}\sigma_i(\nu)\nu^{-1}d\nu}}}
{\displaystyle{\int_{\nu_0}^{~n+1}}{B_{\nu}(T_e)\sigma_i(\nu)}\nu^{-1}d\nu},
\end{equation}
where the summation in the numerator is now over $n$ bands, each running over
a frequency range $\nu_j \rightarrow \nu_{j+1}$. In practice, the
monochromatic mean intensity in each band is modelled as either a power law, 
\begin{equation}
J_{\nu,j}=K_{pl}\nu^{\alpha_{pl}},
\end{equation}
or an exponential 
\begin{equation}
J_{\nu,j}=K_{exp}e^{(-h\nu/kT_{exp})}.
\end{equation}
The values for the four parameters $K_{pl}$, $\alpha_{pl}$, $K_{exp}$ and $T_{\exp}$, 
are deduced from the two band-limited radiation field estimators mentioned
in Section~\ref{itscheme}, i.e. $J$ and $\overline{\nu}$. Thus, for
each model, the two free parameters are set by requiring that
integrating $J_{\nu}$ and $\nu J_{\nu}$ over frequency yields the 
correct values for $J$ and $\overline{\nu}$. The choice between the
exponential and power-law models is finally made by comparing the
third band-limited estimator of the radiation field, $\sigma_{\nu}$,
to the value predicted by each model. 

\subsubsection{Heating and cooling}
\label{heat_cool}
Earlier versions of \textsc{python} accounted for heating
and cooling due to free-free, free-bound and bound-bound processes, as
well as for adiabatic cooling due to the expansion of the
wind. For this study, we have added Compton processes and dielectronic
recombinations to the heating and cooling processes included in the
code. The former, especially, can be important in QSO winds.

\paragraph{Compton Heating and Cooling}

Following \cite{sim_2010}, we calculate the Compton heating rate in a cell with electron density $n_e$ as
\begin{equation}
H_{comp}=n_e\sum\bar{f}(\nu)\sigma_C(\nu)W\, ds,
\end{equation}
where the summation is carried out over all photon packet paths of
length $ds$ in the cell. Each photon packet carries luminosity (weight) $W$
$\rm{ergs~s^{-1}}$ and has frequency $\nu$.  Here,  $\bar{f}(\nu)$  is
the mean energy lost per interaction, equal to $h\nu/m_ec^2$ 
averaged over scattering angles. The Klein-Nishina formula is used to
compute the cross section $\sigma_C$.

We also include induced Compton heating in our calculations. Following
\cite{hazy3}, we estimate this heating rate as    
\begin{equation}
H_{ind~comp}=n_e\sum\eta_i(\nu)\bar{f}(\nu)\sigma_C(\nu)W\,ds,
\end{equation}
where $\eta_i(\nu)$ is the photon occupation number given by
\begin{equation}
\eta_i(\nu)=\frac{J_{\nu}}{\frac{2h\nu^3}{c^2}}.
\end{equation}

Finally, the Compton cooling rate ($C_{comp}$) is given by
\begin{equation}
C_{comp}=16\pi V\sigma_TJ\frac{kT_e}{m_ec^2},
\end{equation}
where $V$ is the volume of the wind cell.

\paragraph{Dielectronic Recombination Cooling}

At the high temperatures expected in a wind
irradiated by X-ray photons, dielectronic recombinations can become
a significant recombination channel. Even though the associated
energy loss from the plasma is never a major cooling process, we
include it in our calculations to maintain internal consistency
(i.e. processes affecting both ionization and thermal equilibrium
should be represented in both calculations).
We estimate the cooling rate due to dielectronic recombinations as
\begin{equation}
C_{DR}=Vn_e kT_e \sum_{All\ ions}{{n_i}{\alpha_{DR}^{i}(T_e)}},
\end{equation} 
where $\alpha_{DR}^{i}(T_e)$ is the temperature-dependent dielectronic recombination rate
coefficient for each ion in a cell. The main approximation made in
this equation is that the energy removed from the electron pool per
dielectronic recombination is equal to the mean kinetic energy of an
electron ($\simeq kT_e$).

\subsubsection{Code Validation}

In order to validate the design and implementation of our ionization
algorithm, we have carried out a series
of tests against the well-known photoionization code \textsc{cloudy}
\citep{cloudy}. In these tests, we consider a geometrically thin
spherical shell illuminated by a wide range of input SEDs, including
specifically SEDs with significant X-ray power-law components. We
generally find good agreement between the ionization states predicted by 
\textsc{python} and \textsc{cloudy} in these tests for species
important in the BAL context. We also find good agreement in the
predicted electron temperatures for quite a wide range of models.  

\begin{figure}
\plotone{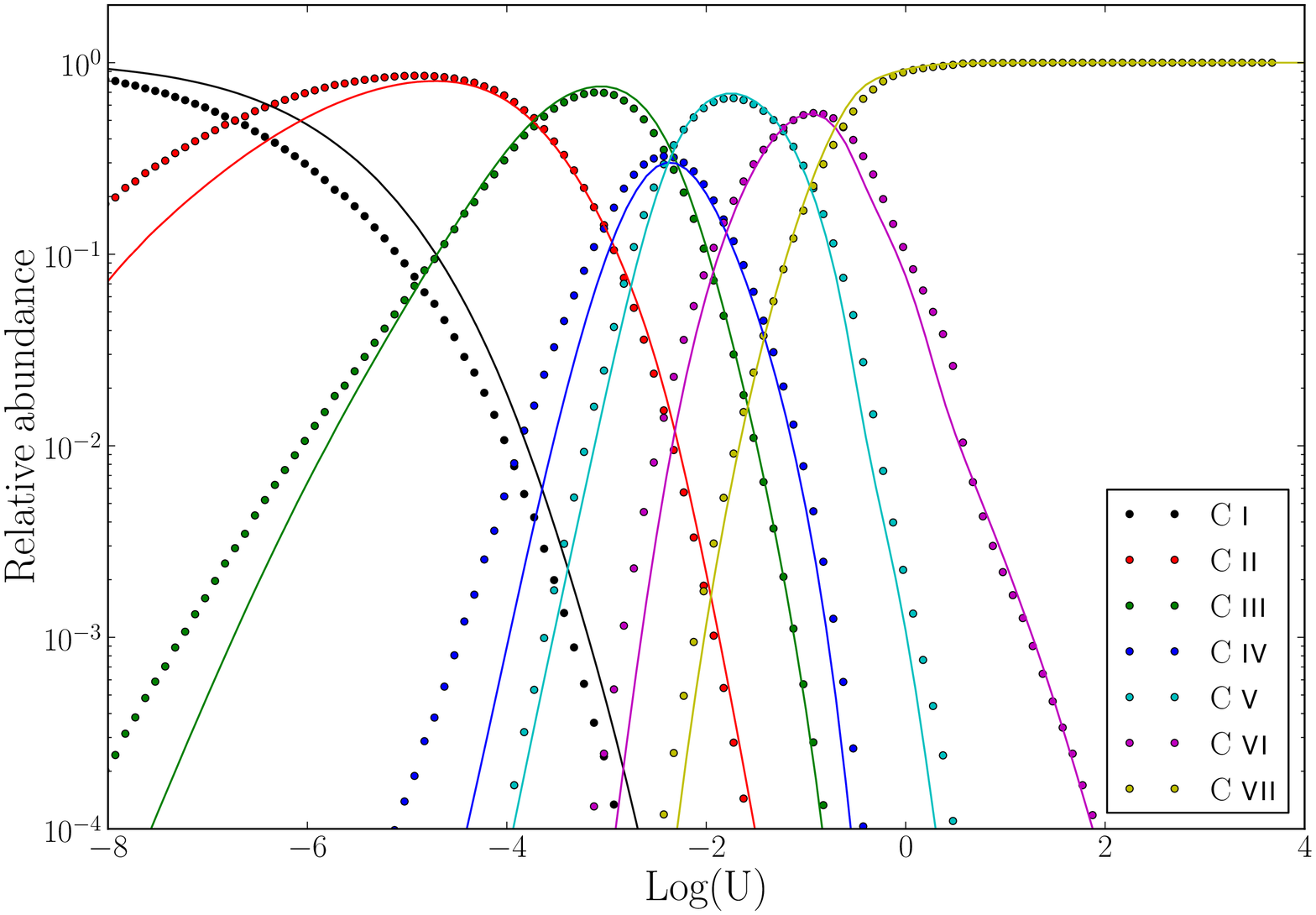}
\caption{Relative abundance of different carbon ionization stages for
  a range of ionization parameters U. Calculations carried out in a
  thin shell geometry illuminated by a broken power law. Circles are
  predictions from \textsc{python} and lines are predictions from
  \textsc{cloudy}.}
\label{carbon_IP}
\end{figure}

One example of these tests is shown in Figure~\ref{carbon_IP}. Here, we plot
the relative abundances of the various ionization stages of carbon as
a function of ionization parameter
\begin{eqnarray}
U&=&\frac{Q_H}{4\pi R^2 n_{H} c} \label{IP_theory}\\ 
&=& \frac{\displaystyle{\int_{13.6{\rm eV}}^{\infty} } (L_\nu/h\nu) d\nu}{4\pi R^2  n_Hc} \\
  &=& \frac{4\pi}{n_Hc}\int_{13.6{\rm{eV}}}^{\infty}\frac{{J_{\nu}d\nu}}{h\nu}
\end{eqnarray}
where $\rm{Q_H}$ is the number of hydrogen-ionizing photons emitted by
the illuminating source per second, $R$ is the distance to the source,
$\rm{n_H}$ is the local hydrogen number density and $c$ is the speed of
light. $L_\nu$ is the monochromatic luminosity.  
The ionization parameter is a measure of the ratio of the ionizing photon
density and the local matter density. As such, it is a good
predictor of the ionization state of optically thin photoionized
plasmas.

In the test shown in Figure~\ref{carbon_IP}, the irradiating SED was
modelled as a a doubly-broken power law with $\alpha = 2.5$ below
$0.136$~eV, $\alpha = -2$ above $20$~keV and $\alpha = -0.9$ 
between these break frequencies. This was in order to allow direct 
comparison with a \textsc{cloudy} model defined via the \texttt{power law}
command. The spherical shell was assumed to lie at $R = 10^{11}$~cm, the
hydrogen density was taken to be $n_H = 10^7 {\rm ~cm^{-3}}$ and the
ionization parameter was adjusted by varying the luminosity of the ionizing 
SED. With such a simple set-up, both the SED and the geometry can be
modelled in exactly the same way in both \textsc{python} and
\textsc{cloudy}. Given the likely differences in detailed 
atomic physics and atomic data used in \textsc{cloudy} and \textsc{python}, there is good 
agreement between the codes, especially for the moderately to
highly ionized stages of carbon that we expect to see in the vicinity of 
a QSO. This agreement is illustrated in Figure \ref{carbon_IP}.

As noted above, the electron temperatures predicted by \textsc{python} and
\textsc{cloudy} also generally agree quite well. One exception to this
is situations in which iron lines dominate the cooling, where \textsc{python} 
can overestimate the temperature by as much as a factor of about
2. Some wind regions in our benchmark model are likely to be affected
by this. However, since the electron temperature only appears in the
ionization equation via $\sqrt{T_e}$, the effect of this issue on the
relevant ionization fractions in our models is relatively small.

\subsection{Spectral Synthesis}
\label{spec_synth}

Once the thermal and ionization state of the wind has converged, the
ionization cycles are terminated and the spectral cycles begin. In the
latter, the thermal and ionization structure are considered fixed, and
photons are only generated over the restricted frequency range for
which detailed predictions are needed. This saves CPU time and allows us
to achieve much greater spectral resolution and signal-to-noise in the
simulated spectra.

The spectra themselves are extracted from the Monte Carlo simulation
using the techniques described by LK02. Collisionally excited line
emission is treated in the two-level atom approximation, which is
acceptable for strong resonance 
lines in which the upper level is populated by transitions from the
lower (ground) state. Most of the transitions associated with BALs 
and BELs fall into this category. However, this approximation is {\em
not} adequate for many other transitions, most notably the Hydrogen Lyman
and Balmer series, for which the upper states are populated by a
radiative recombination cascade. \citep{sim_drew_05} have already used 
a more general method to model infrared H{\sc i} line emission in YSOs
with \textsc {python}; work is in progress on implementing this 
treatment into our simulations of AGN/QSOs.

\subsection{Atomic data}
\label{atomic_dat}

For the calculations described in this paper, we include atomic data
for H, He, C, N, O, Ne, Na, Mg, Al, Si, S, Ar, 
Ca and Fe. Elemental abundances are taken from \cite{vbt} and the
basic atomic data, such as
ionization potential and ground state multiplicities, are taken from
\cite{vvf}. For H, He, C, N, O and Fe, we use \textsc{topbase}
\citep{cmoz} data for energy levels and photoionization cross sections
from both ground state and excited levels. For other elements, we use
the analytic approximations for ground state photoionization cross sections
given in \cite{vfky}. For bound-bound transitions, we use the line
list given in \cite{vvf}, which contains nearly 6000
ground-state-connected lines. The levels associated with these lines
are computed from the line list information using a method similar to
that described by \cite{lucy}. 

We adopt dielectronic recombination rate coefficients and total radiative
rate coefficients from the \textsc{chianti} database version 7.0
\citep{dere,landi}. Ground state recombination rates are taken from
\cite{badnell} where available and otherwise computed from the photoionization
data via the Milne relation. Finally, we use Gaunt factors for 
free-free interactions given by \cite{sutherland}.

\section{A Benchmark Disk Wind Model For BALQSOs}
\label{fid_mod} 

One of the main challenges in trying to model BALQSOs is that the
physical characteristics of the disk winds that produce the BALs are
highly uncertain. The kinematic model described in Section~\ref{kin_mod}
is sufficiently flexible to describe a wide range of outflow
conditions, but the price for this flexibility is a potentially huge
parameter space. We have therefore used a variety of theoretical and
observational constraints (and considerable trial and error) to design
a benchmark disk wind model whose predicted spectrum resembles that of
a BALQSO (for sightlines cutting through the outflow). 

Since we have not yet carried out a systematic exploration of the
relevant parameter space, we make no claim that this fiducial model
is optimal, nor that it faithfully describes reality. Indeed, in
Section~\ref{discussion}, we will discuss some of its
shortcomings. However, we believe it is nevertheless extremely useful as a
benchmark. More specifically, it provides (i) a first indication of
what can be realistically expected of disk wind models; (ii) a better
understanding of the challenges such models face; (iii) a convenient
starting point for further explorations of parameter space. 

In the rest of this section, we will discuss the constraints that 
have informed our choice of parameters for the benchmark model, which
are summarised in Table \ref{wind_param}. 

\subsection{Basic Requirements: BALnicity and Ionization State}

Our primary aim here is to find a set of plausible parameters for a
QSO wind that yield  synthetic spectra containing
 the features expected of a BALQSO from a range of viewing
angles. As  noted in Section~\ref{introduction}, in the majority of
BALs, the HiBALs, the absorption
features are due to highly ionized species such as N~{\sc v}
$\lambda\lambda$1240, C~{\sc iv} $\lambda\lambda$1550, and Si~{\sc iv}
$\lambda\lambda$1400.  The C~{\sc iv} feature, in particular, is most often
used to identify BALQSOs, so we judge the initial success of candidate
wind models by their ability to produce this feature. 
In practice, this is primarily a constraint on the ionization state of
the wind: we require that C~{\sc iv} should be present -- if not
necessarily dominant -- throughout an appreciable fraction of the
outflow.

BALs are usually identified via the so-called BALnicity index (BI)
\citep{weyman}, which is a measure of absorption strength in the
velocity range between $-3000\rm{~km~s}^{-1}$ and
$-25000\rm{~km~s}^{-1}$. It is therefore not sufficient to produce
C~\textsc{iv} anywhere in the outflow -- in order for the 
model to resemble a BAL, C~\textsc{iv} needs to be present in
reasonable concentrations in regions characterized by the correct
velocities. 

\subsection{Black Hole Mass, Accretion Rate and Luminosity}

Our goal is to model a fairly typical, high-luminosity (BAL)QSO, so we
adopt a black hole mass of $10^9~\rm{M_{\odot}}$, along with an Eddington
ratio of $\epsilon = 0.2$. For an assumed accretion efficiency of
$\eta=0.1$, this corresponds to an accretion rate of $\rm{\dot{M}_{accretion}}\sim
5~\rm{M_{\odot} yr^{-1}}$ and a bolometric disk luminosity of $L_{bol}
\simeq 2.4\times 10^{46}~\rm{ergs~s^{-1}}$. 

\subsection{Outflow Geometry}

 BALQSOs represent $\simeq\!20\%$ of all QSOs (e.g. \citealt{knigge_balfrac, hew_foltz}, but also
see \citealt{allen_2011}). There are two obvious ways to interpret this
number: (i) either all QSOs spend $\simeq\!20\%$ of their life as BALQSOs
(``evolutionary unification''; e.g. \citealt{becker_2000}), or (ii) all
QSOs contain disk winds, but the resulting BAL signatures are visible
only from $\simeq 20\%$ of all possible observer orientations (``geometric
unification''; e.g. \citealt{elvis_2000}).\footnote{This could be an underestimate, since intrinsic BALQSO
fractions tend to be derived from magnitude-limited samples, with no
allowance for the possibility that foreshortening, limb-darkening or
wind attenuation may reduce the continuum flux of a typical BALQSO
relative to that of a typical QSO \citep{krolik_voit}. For the purpose of the
present paper, we ignore this potential complication.} Here, we adopt the geometric
unification picture, with the aim of testing its viability.

\begin{figure}
\plotone{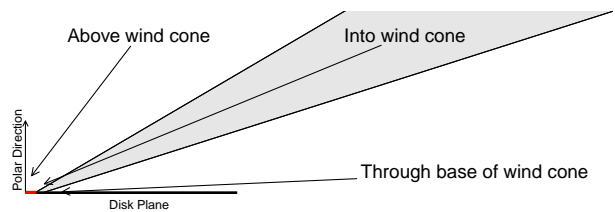}
\caption{Cartoon illustrating the three broad classes of sightline discussed in the text.}
\label{sightlines}
\end{figure}

As illustrated in Figure~\ref{sightlines}, our wind geometry allows
for three distinct classes of sightline towards the central engine. Looking from the polar direction,
``above the wind cone'', the sightline does not intersect the wind at all, and looking from the equatorial direction,
the sightline passes through the base of the wind cone  where the projected line of sight
velocities are too low to produce BAL features. We anticipate that BAL features will result when the central engine
(or, strictly speaking, 
the UV-bright portions of the accretion disk) are viewed from the third class of sightline - looking
``into the wind cone''. We therefore 
expect such sightlines to subtend $\simeq$~20\% of the
sky as seen by the central engine. Note that this description of the sightlines only applies
to models like our benchmark model, in which most of the emission arises from
inside the wind launching radius.

If we define $\theta_{min}$ and $\theta_{max}$
as the angles made by the inner and outer boundaries of the
wind with respect to the disk normal, these constraints translate into
a simple relationship between $\theta_{min}$ and $\theta_{max}$, i.e. 
\begin{equation}
0.2=\cos(\theta_{min})-\cos(\theta_{max}).
\end{equation}

Most geometrical unification scenarios assume that BAL-producing
QSO winds flow predominantly along the equatorial plane of the disk,
rather than along the polar axis (e.g. \citealt{weyman,ogle_99,elvis_2000}; 
however, see \citealt{zhou_06} and \citealt{brotherton_06} for a contrary view).
In line with this general view, we have
found that ionization states capable of producing C~\textsc{iv} and
other typical BAL features are
more naturally produced in equatorial, rather than polar disk
winds. This is because one of the main challenges faced by disk wind
models is to prevent
the over-ionization of the outflow by the central X-ray source (see
below). Other things being equal, equatorial winds are less ionized, partly because of
limb darkening, and partly because the disk appears foreshortened from equatorial
directions. In addition, in geometries like ours, where 
the wind arises from parts of the disk radially outside the
location where the ionizing radiation is produced, an equatorial wind exhibits 
self-shielding, i.e. the radiation must pass through the inner parts of the wind 
to reach the outer parts. Therefore the ionization state can be lower than for a 
polar wind, where all parts of the wind can see the ionizing radiation directly.

Based on these considerations, we adopt a mostly
equatorial geometry with $\theta_{min}=70^{\circ}$ and $\theta_{max}=82^{\circ}$ for our
benchmark model. 
With this choice, 66\% of the remaining sightlines lie ``above'' the wind cone, while 14\%
view the central source through the base of the wind. If geometric unification is correct and
complete -- in the sense that the emission lines in non-BAL QSOs are
produced by the same outflow that produces the absorption features
seen in BALQSOs \citep[e.g.][]{elvis_2000} -- we would ideally see different
classes of QSO as we view a given system from the three distinct types
of viewing angles.

\subsection{Wind-Launching Region}
\label{wind_launch}

There is no consensus on the location of the wind-launching region on
the accretion disk. However, there are several empirical and
theoretical considerations that can inform the design of our benchmark
model. 

First, within the geometric unification scenario, QSO disk winds are
responsible not only for BAL features, 
but also for the broad emission lines (BELs) seen in non-BAL QSOs. We
can therefore use the result of reverberation mapping studies of this
broad line region (BLR) to guide us in locating the
outflow. \cite{kaspi} have carried out reverberation mapping of the
C~\textsc{iv} BLR for high luminosity quasars. A representative
example is S5 0836+71, a quasar with a black hole mass of
$2.6\times10^{9}~M_{\odot}$, for which they found a rest frame delay of
188 days. This implies a distance of the line-forming region from the
UV continuum source of $1260~\rm{r_g}$.\footnote{Note that the UV continuum
is produced at very small radii compared to the wind
launch radius and so can be treated as originating from the centre of
the model for these estimates}
Since the C~\textsc{iv} line-forming region may lie substantially
downstream in the outflow, this is an upper limit on the distance of
the launching region from the centre.

Second, BAL features exhibit variability on timescales from days to
years \citep[e.g.][]{capellupo_11,capellupo_12,capellupo_13}. If
interpreted in terms of wind features crossing our line of sight, the
{\em shortest} variability time scales -- which arise in the innermost
parts of the line-forming region -- correspond to distances of 
$\sim 3 \times 10^{16}$~cm or, equivalently, $\sim 200~r_g$ from the
centre (for an assumed $M_{BH} \simeq 10^9~M_{\odot}$).    
 
Third, one of the most promising mechanisms for driving QSO disk winds
is via radiation pressure in spectral lines 
\citep[e.g.][]{proga_stone_kallman,proga_kallman}. Empirical support for this idea
comes from the presence of the ``ghost of Lyman $\alpha$'' in some
N~\textsc{v} BAL features 
(see \citealt{arav_95,arav_96,north_knigge_goad}; but also see \citealt{cottis_10}).
In published simulations where such winds are modelled in a physically
consistent manner, the typical launching radii are $\rm {(a~few)} \times 100~r_g$
 \citep[e.g.][]{proga_stone_kallman,ris_elv}.

Fourth and finally, it is physically reasonable to assume that the
maximum velocity achieved by the outflow corresponds roughly to the
fastest escape velocity in the wind launching region on the accretion
disk. Observationally, $v_{max} \sim 20,000~\rm{km~s^{-1}}$, which, for
a $10^9M_{\odot}$ black hole, corresponds to the escape velocity from 
$\simeq\!400~r_g$.

Taken together, all of these considerations suggest that a reasonable
first guess at the location of the wind-launching region on the disk
is on the order of a few hundred $r_g$. In our benchmark model, we
therefore adopt $r_{min} = 300~r_g$ and $r_{max} = 600~r_g$.

\subsection{Wind Mass-Loss Rate}

One of the most fundamental parameters of any wind model is the
mass-loss rate into the outflow, $\dot{M}_{wind}$. In our benchmark
model, we set $\dot{M}_{wind} = \dot{M}_{acc} =
5~M_{\odot}~\rm{yr}^{-1}$. This value was arrived at mainly by trial
and error, but with a conservative preference for values satisfying
$\dot{M}_{wind} \ltappeq \dot{M}_{acc}$. As shown explicitly in 
Section~\ref{xray_sen}, in practice, we found that we required
$\dot{M}_{wind} \simeq \dot{M}_{acc}$ in order to produce BAL features
for even modest X-ray luminosities. 

Strictly speaking, any model with $\dot{M}_{wind} \gtappeq
\dot{M}_{acc}$ is not entirely self-consistent, since the presence of
such an outflow would alter the disk's temperature 
structure \citep[e.g.][]{knigge_99}. However, we neglect this
complication, since the vast majority of the disk luminosity arises
from well within our assumed wind-launching radius. Thus even though our model
ignores that the accretion rate {\em is} higher further out in the
disk, it correctly describes the innermost disk regions that produce
virtually all of the luminosity. 

Finally, in the absence of evidence to the contrary, we take the
simplest possible prescription for the run of the mass-loss rate with
radius across the wind-launching region, i.e. $\lambda = 0$ (see
Equation~\ref{lambda_eqn} in Section~\ref{kin_mod}).

\subsection{Velocity Law Parameters}

Ideally, the parameters defining the poloidal velocity law of the wind
would be predicted by the relevant acceleration mechanism. However,
since this is not yet possible, we have simply adopted parameters that
result in reasonably BAL-like C~\textsc{iv} features.

As noted in Section~\ref{wind_launch} above, it seems physically
plausible that the terminal outflow velocity along a poloidal
streamline will roughly reflect the local escape velocity at the
streamline footpoint. Having already adopted wind launching radii  
based on this idea, we therefore also adopt $v_{\infty} = v_{esc}$,
i.e. $f = 1$ in Equation~\ref{v_infty}.

The other two parameters defining the poloidal velocity law are $R_v$,
the poloidal distance at which the velocity is equal to half 
the terminal velocity, and $\alpha$, the exponent describing how the
velocity varies with poloidal distance. As discussed above, the locations 
of BAL- and BEL-forming regions within the wind likely extend to
distances of up to about $2000~r_g$, so this is a reasonable number to
adopt for $R_v$. For the $10^9 M_{\odot}$ black hole in our benchmark
model, this corresponds to $R_v = 10^{18}~$cm. Finally, in the absence
of other information, we adopt $\alpha=1$. This produces a relatively
slow acceleration as a function  of poloidal distance along a
streamline and seems to produce reasonable BAL shapes.
We will explore the effect of changing these parameters in future
work.

\subsection{X-ray Spectrum and Luminosity}
\label{xray_spec_lum}

The X-ray luminosity of the central source is set to
$\rm{L_{X}=10^{43}~{\rm ergs~s^{-1}}}$ in our benchmark model. This
value was arrived at through the requirement that
the model should produce BALs which constrains the X-ray
luminosity, since X-rays can strongly affect the ionization state of
the outflow. We set the spectral index of the X-ray spectrum to
$\alpha_X = -0.9$, as suggested by \cite{giustini_08}.

Quasar spectra are often characterised in terms of the ratio of 
X-ray to bolometric (or UV) luminosity. This ratio
is usually characterized by the spectral index between X-rays and
optical/UV, $\alpha_{OX}$, defined as 
\begin{equation}
\alpha_{OX}=0.3838\log\left(\frac{L_{\nu}(2~keV)}{L_{\nu}(2500~\mbox{\scriptsize{\AA}})}\right),
\end{equation}
where $L_{\nu}(2~keV)$ and $L_{\nu}(2500~{\AA})$ are the inferred
monochromatic luminosities at 2~keV (X-ray) and 2500~\AA (UV),
respectively. Ignoring the angle dependence of the disk emission, we
obtain a mean value of $\alpha_{OX}\sim-2.2$ for the parameters used
in our benchmark model. This is considerably lower than observed in
typical non-BAL QSOs \citep{just_07} and lies near the lower (X-ray
weak) end of the range of $\alpha_{ox}$ seen in BALQSOs
\citep[e.g][]{giustini_08,gibson_09}.  The X-ray weakness of the
model, as well as the sensitivity of our results to the adopted X-ray
luminosity, are discussed in more detail in Section~\ref{xray_sen}.

\subsection{Size and Resolution of the Numerical Grid}

The outer edge of our simulation grid is set to
$1\times10^{19}\rm{~cm}$. This is large enough to ensure that all
spectral features of interest arise entirely within the computational
domain. In all simulations shown in this paper, this domain is sampled
by a cylindrical grid composed of $100 \times 100$ logarithmically
spaced cells. We have carried out resolution tests to ensure that this
is sufficient to resolve the spatially varying physical conditions and
ionization state of the outflow.

\begin{table}
\begin{tabular}{p{3cm}p{4cm}}
\hline Free Parameters 	&	 Value \\ 
\hline \hline 
$M_{BH}$ 	 &	 $1\times 10^9~\rm{M_{\odot}}$ \\ 
$\dot{M}_{acc}$ 	 &	 $5~M_{\odot}yr^{-1} \simeq 0.2~\dot{M}_{Edd}$\\ 
$\alpha_X$ 	 &	 $-0.9$ \\ 
$L_{X} $ 	 &	 $1\times10^{43}~\rm{ergs~s^{-1}}$\\ 
$r_{disk}(min)=r_{X}$   &	 $6r_g=8.8\times10^{14}~{\rm cm}$ \\ 
$r_{disk}(max)$   &	 $3400r_g = 5\times10^{17}~{\rm cm}$ \\ 
$\dot{M}_{wind}$  &	 $5~M_{\odot}yr^{-1}$ \\ 
$r_{min}$ 	&	 $300r_{g} = 4.4\times10^{16}~{\rm cm}$\\ 
$r_{max}$ 	&	 $600r_{g} = 8.8\times10^{16}~{\rm cm}$ \\ 
$\theta_{min}$ 	&	 $70.0^{\circ}$ \\ 
$\theta_{max}$ 	&	 $82.0^{\circ}$ \\ 
$\lambda$ 	&	 $0$ \\ 
$v_{\infty}$ 	&	 $v_{esc}$(f=1) \\ 
$R_v$ 	        &	 $1\times10^{18}$cm \\ 
$\alpha$ 	&	 $1.0$ \\
\hline 
Derived Parameters 	&	 Value \\ 
\hline \hline
$L_{\nu}(2500\mbox{\scriptsize{\AA}})$&	 $6.3\times10^{30}~\rm{ergs~s^{-1}~Hz^{-1}}$\\ 
$L_{\nu}(2keV)$   &	 $1.2\times10^{25}~\rm{ergs~s^{-1}~Hz^{-1}}$\\ 
$L_{bol}$ 	 &	 $2.4\times10^{46}~\rm{ergs~s^{-1}}$\\
$M_{bol}$ 	 &	 -27.4\\ 
$M_u$            &	 -26.2\\ 
 $\alpha_{OX} $ 	 &	 -2.2\\ 
\end{tabular}
\caption{Wind geometry parameters used in the benchmark model.}
\label{wind_param}
\end{table}

\section{Results and Analysis}
\label{results}

Having defined a benchmark disk wind model for AGN/QSOs, we are ready
to examine its predicted physical and observable
characteristics. In Section \ref{synth_spectra} we will present synthetic spectra
reminiscent of BALQSOs for sightlines looking into the wind cone. 
However, it is helpful to examine the physical and ionization
state of the benchmark model before analyzing these spectra in
detail.

\subsection{The Physical State of the Outflow}

\begin{figure*}
\plotone{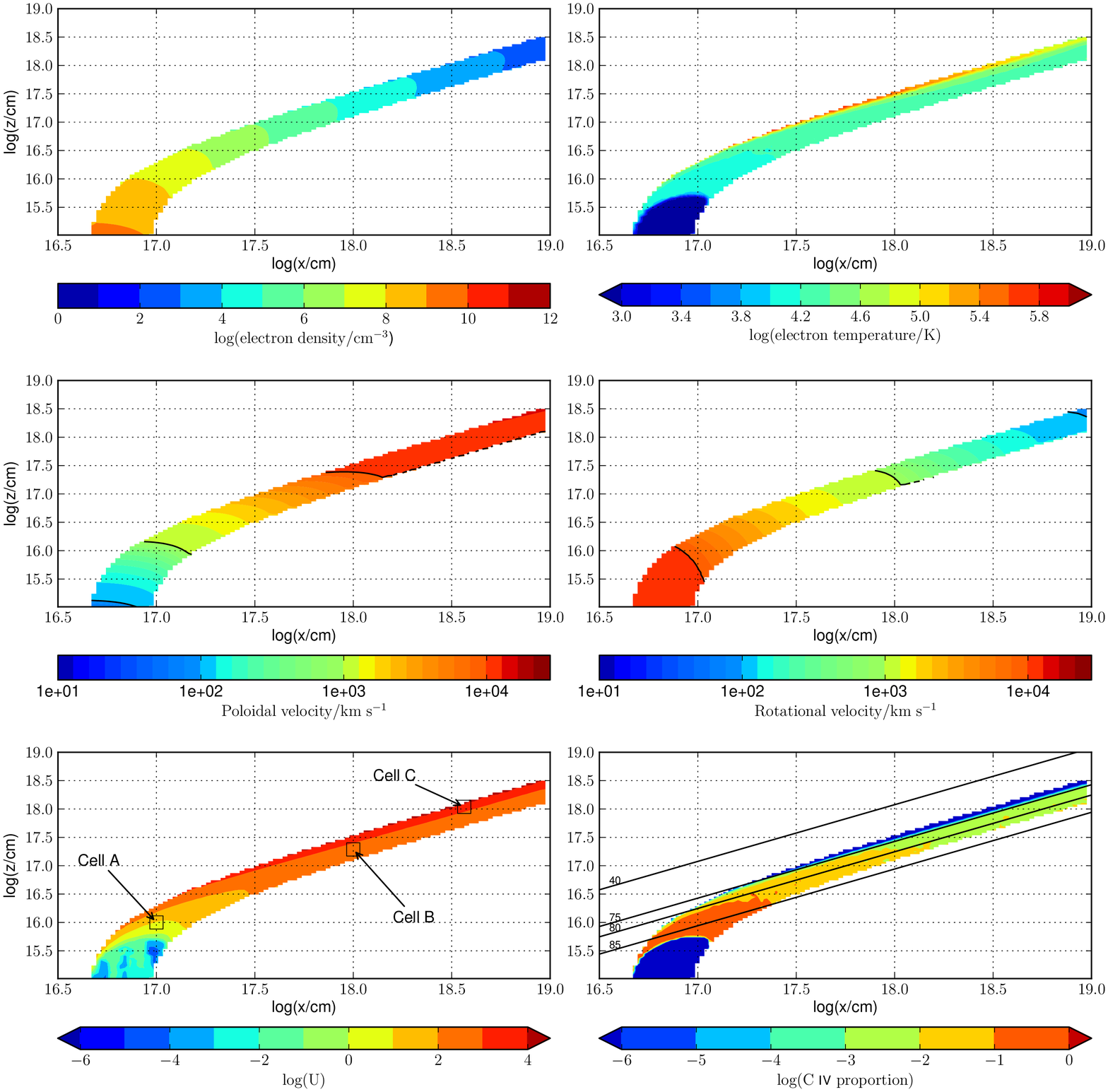}
\caption{The electron density (top left), electron temperature (top
  right), poloidal velocity (mid left), rotational velocity (mid
  right), ionization parameter (bottom left) and
  proportion of C ions in the C~\textsc{iv} ionization state
  (bottom right) for the
  benchmark model. Only the positive xz plane is shown, the wind is
  rotationally symmetrical around the z axis. Note the logarithmic
  scales and the difference in scales for the x and z axes. The location 
  of cells for which `cell' spectra are presented in figure \ref{cell_spectra} are
 shown in the ionization parameter plot and the black
  lines on the C~\textsc{iv} plot show sightlines through the wind to
  the origin used to produce the spectra plotted in figure
  \ref{spectra}.}  
\label{physical_wind}
\end{figure*}

The top four panels in Figure~\ref{physical_wind} show a selection of
physical parameters of the converged wind model. Considering first the
electron density (top left panel), we see that our choice of kinematic
parameters has given rise to an outflow with a fairly high density of
$n_H \sim 10^{10}~\rm{cm^{-3}}$ at its base. However, this declines quickly
as we move outwards, due to the expansion and acceleration of the
wind. Hydrogen is fully ionized throughout the entire
outflow, so $n_H \simeq n_e$ everywhere. 

The electron temperature in the wind ranges from 
$\sim 10^3~{K}$ near the base of the wind to more than $10^5~{K}$ (top 
right panel). The highest temperatures are  in a thin layer near the ``top'' of the wind, at distance of
$\simeq\!10^{18}~\rm{cm}$ from the central engine. These regions are
hot because they are directly exposed to the radiation field of the
accretion disk and the X-ray source. These regions also shield the wind material
``behind'' them, however, and thus help to ensure more moderate
temperatures in the rest of the outflow. In fact, much of the disk
wind is heated to temperatures near $T_e \sim 10^4~{\rm K}$, which are
quite conducive to the formation of the high-ionization lines
typically seen in (BAL)QSOs, such as C~\textsc{iv}, Si~\textsc{iv} and
N~\textsc{v}.

The middle panels of Figure~\ref{physical_wind} illustrate the velocity
structure of our benchmark model, separated into poloidal
(middle left panel) and rotational (middle right panel) components. 
The poloidal velocities show the relatively gradual
speed-up of the outflow as a function of distance, which is due to our 
choice of velocity law exponent. As one would expect, low  
velocities are found near the base of the wind and high velocities are reached
only quite far out. It is this variation in poloidal velocity which produces
BAL features as photons produced by the effectively point-like central
UV source are scattered out of the line of sight by progressively faster
moving wind parcels.
 
By contrast, the highest rotational velocities are found 
near the disk plane, where they are effectively equal to the Keplerian
velocities in the disk. They then decline linearly with increasing
cylindrical distance from the rotation axis, because we assume that wind material
conserves specific angular momentum. The projected rotational velocity
along a sightline to the central source is zero (or nearly so), and therefore
rotational velocities do not play a major part in the formation of BAL
features, although they should help in producing broad {\em emission} lines.

\begin{figure}
\plotone{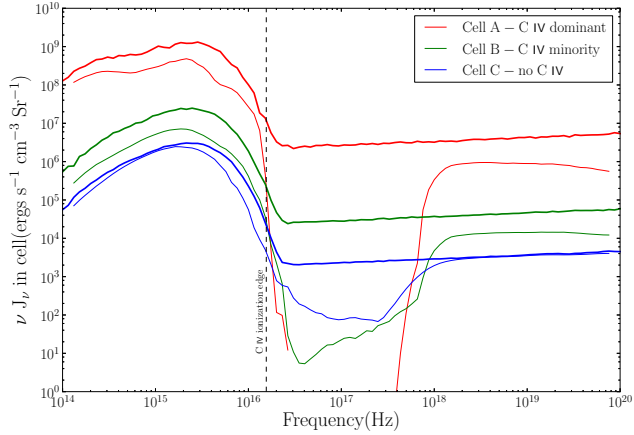}
\caption{Mean specific intensity in three cells. The red line is a cell with
  C~\textsc{iv} as the dominant Carbon ionization stage, in the base
  of the wind. The green line is for a cell with significant
  C~\textsc{iv} in a fast moving part of the wind, and the blue line is
  for a cell at the top of the wind. fully exposed to the ionizing
  continuum. The thick lines represent the unobscured spectrum 
  that would be
  seen in that region. The vertical line
  marks the location of the C~\textsc{iv} photoionization edge.}
\label{cell_spectra}
\end{figure}

The bottom left and right panels in Figure~\ref{physical_wind} show
the ionization parameter and C~\textsc{iv} ionization fraction
throughout the outflow, respectively. Interestingly, the model
generates sufficient concentrations C~\textsc{iv} to make strong BALs
even though $U > 1$ throughout most of the wind. This contrast with
the results shown for a thin-shell model in Figure \ref{carbon_IP},
where this ion is not present for such values of $U$. The main reason for
this difference is that the wind is self-shielding: absorption
in the inner parts of the outflow (primarily due to photoionization)
can strongly modify the SED seen by the outer parts of the wind. This,
in turn, changes the ionization parameter at which particular ions are
preferentially formed. Figure \ref{cell_spectra} illustrates 
the effect of the self-shielding on the SED seen by different parts of
the outflow.

\subsection{Observing the Model: Synthetic Spectra}
\label{synth_spectra}

\begin{figure*}
\plotone{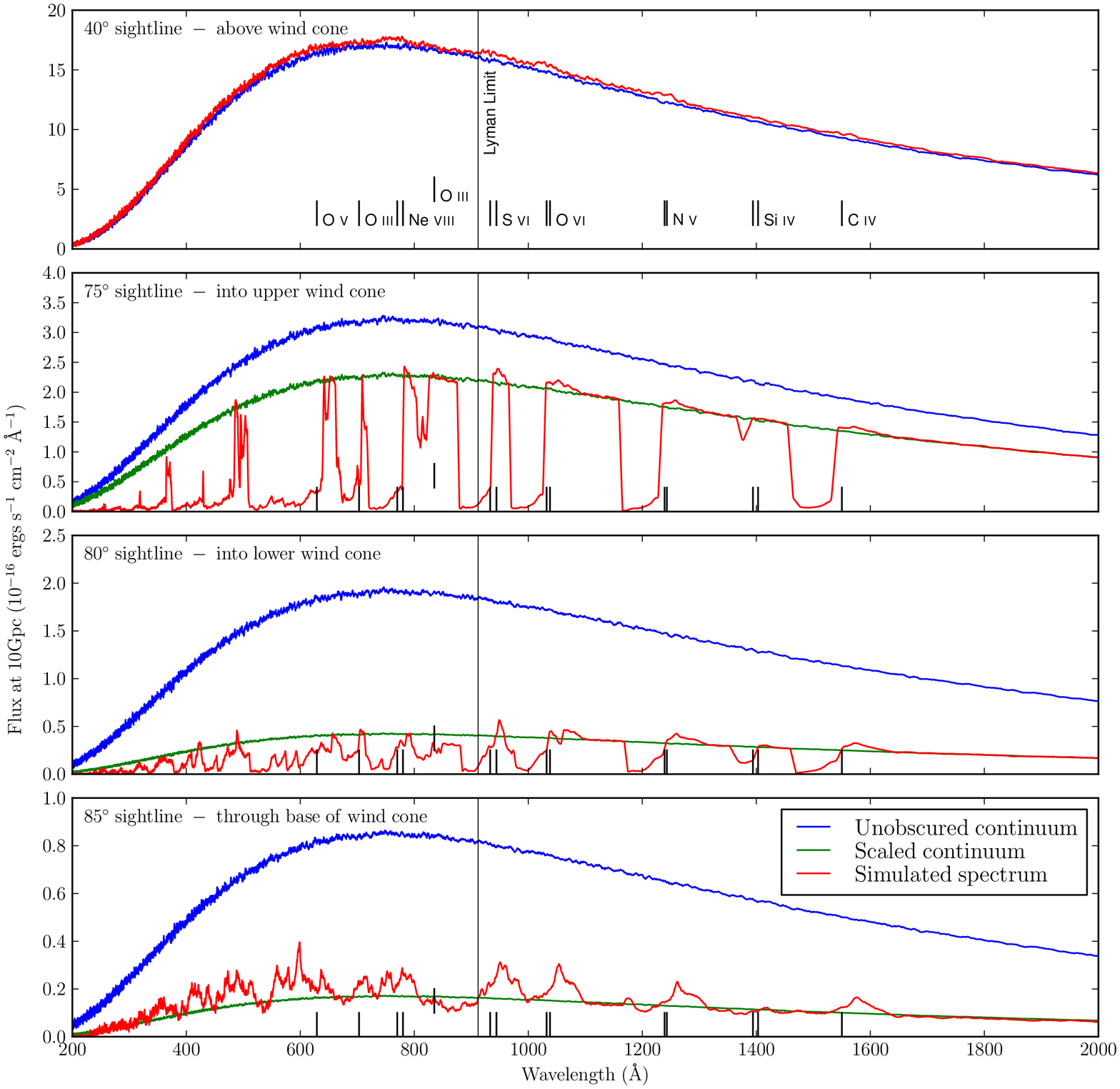}
\caption{Simulated spectra for four sightlines.  
The top panel shows the
  $40^{\circ}$ sightline, over the top of the wind to the brightest
  parts of the accretion disk, whilst the next panel shows the
  $75^{\circ}$ sightline, looking through the upper parts of the
  wind. The next panel is for $80^{\circ}$ looking through the lower
  part of the wind and finally, the bottom panel shows the $85^{\circ}$
  sightline which is almost equatorial, and views the bright central
  source through the very base of the wind. For each sightline, the unobscured
  continuum is also plotted for comparison along with a scaled continuum.
  The scaled continuum is the unobscured continuum scaled to equal the
  simulated spectrum away from line features, and therefore takes account of
  electron scattering. The location of some of
  the spectral features most relevant for BALQSOs are marked.
}
\label{spectra}
\end{figure*}

Figure~\ref{spectra} shows the spectra we predict for observers
located along the directions indicated in the lower right panel of
Figure~\ref{physical_wind}. These viewing angles 
correspond to sightlines to the central source that 
look above the wind cone ($i = 40^{\circ}$), into the wind cone 
($i = 75^{\circ}$ and $i = 80^{\circ}$), and through the base of the
wind cone ($i = 85^{\circ}$). For comparison, we also show a pure,
unobscured continuum spectrum for each viewing angle; these continua
were generated by re-running the model with the wind density set to a 
negligible value. 

\subsubsection{Sightlines looking into the wind cone}
\label{into_cone}

The two middle panels in Figure~\ref{spectra} show the emergent
spectra for sightlines through the wind ($i=75^\circ$~and
$i=80^\circ$). These spectra clearly contain BALs, i.e. broad,
blue-shifted absorption features associated with several strong,
high-ionization lines in the UV region. More formally, we calculate
BIs of $11400~\rm{km~s}^{-1}$ ($i=75^\circ$) and $9900~\rm{km~s}^{-1}$
($i=80^\circ$) for the C~\textsc{iv} lines we predict for these two
sightlines. Thus observed systems displaying these spectra would
certainly be classified as bona-fide BALQSOs. This is one of the main
results of our work. 

Many of the transitions exhibiting BAL features in the spectra for both
sightlines also show a redshifted emission component, forming the other
part of the classic P-Cygni profile seen in such sources. This can be
interpreted as the red-shifted part of a classic rotationally broadened,
double-peaked emission line, which is produced primarily by scattering
in the base of the wind. The blue-shifted part of the line profile is
not visible, since it is superposed on and/or absorbed by the BAL
feature. 
  
Another interesting feature of the BAL spectra is that the Si~\textsc{iv} 
absorption feature is narrower than the C~\textsc{iv} feature for all
sightlines. This is due to the lower ionization potential of the silicon ion, 
meaning that it is produced in a more limited part of the wind. The
relative strengths of features is broadly in agreement with 
observations \citep{gibson_09} and demonstrates that, at least to first order,
we are predicting the correct ionization state in the BAL-forming
portion of the wind.  

The continuum for both sightlines through the wind is suppressed 
relative to the unobscured SED (see Figure~\ref{spectra}). 
This attenuation of the continuum is predominantly
due to electron scattering. The optical depth to electron scattering 
through the wind is shown in Figure~\ref{thompson_tau}, along with
the corresponding attenuation factor, $F/F_0 = e^{-\tau_{es}}$.
For example, at $i = 75^{\circ}$ our model gives $\tau_{es} \simeq 0.3$ and
$F/F_0 \simeq 0.7$, in agreement with Figure~\ref{spectra}. 

\begin{figure*}
\plotone{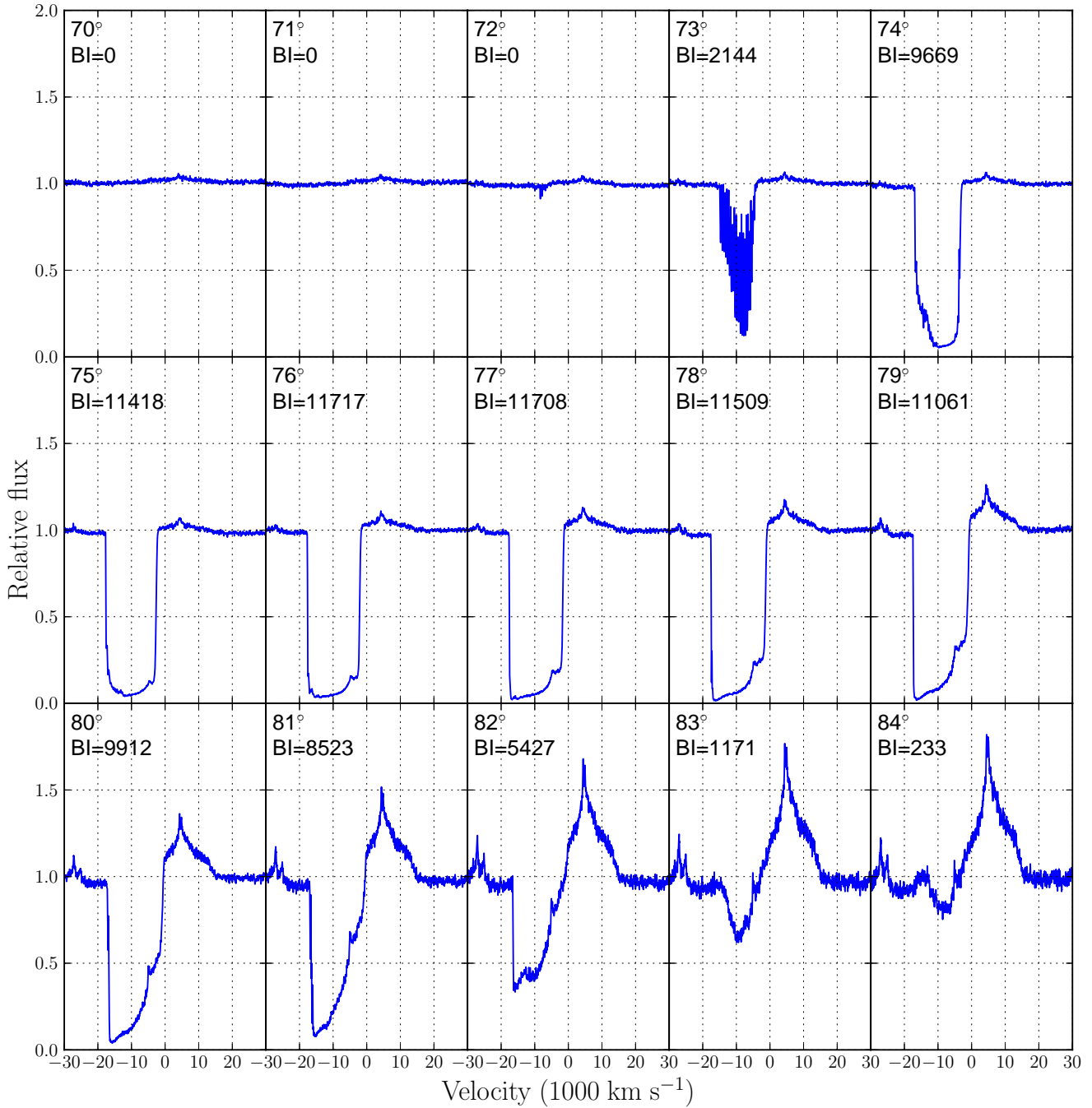} 
\caption{C~\textsc{iv} line detail for sightlines between $70^{\circ}$
  and $84^{\circ}$. Each spectrum is scaled to a linear continuum,
  fitted at velocity $\rm{\pm 30000kms^{-1}}$
The Balnicity Index for the spectrum is marked on each plot.}
\label{detail}
\end{figure*}

Detailed C~\textsc{iv} line profiles predicted by our model for a
more finely spaced grid of sightlines are shown in
Figure~\ref{detail}, along with the $BI$ calculated for each
profile. BALs (i.e. features with $BI > 0$) are observed for
inclinations $73^{\circ} < i < 83^{\circ}$, which represents $\simeq
17\%$ of all possible sightlines.

\subsubsection{Sightlines looking above the wind cone}
\label{above_cone}

We now consider the $i = 40^\circ$ sightline, which views the central
engine from ``above'' the wind cone. In a pure geometric unification
scenario, this sightline may be expected to produce a classic Type I
QSO spectrum. What we observe from the simulation is a slightly
enhanced continuum, along with some broad, but weak emission lines (as
in \cite{sim_2}). 

The continuum enhancement is mainly due to electron scattering {\em
into} this line of sight. As we 
have seen in Section~\ref{into_cone} and Figure~\ref{thompson_tau}, 
the base of the wind is marginally optically thick to electron
scattering in directions along the disk plane. Thus photons
scattering in this region tend to emerge preferentially along the more
transparent sightlines perpendicular to this plane, and the wind
essentially acts as a reflector. 

More importantly, the emission lines superposed on the continuum do 
correspond to the typical transitions seen in Type I QSOs, but they
are weaker than the observed features. For example, 
the C~\textsc{iv} emission line in our model has an equivalent width
of only 1.4~\AA. By 
contrast, the equivalent with of the C~\textsc{iv} line in a typical QSO with a continuum luminosity of
$L_{\lambda}(1550\mbox{\scriptsize{\AA}}) \simeq 10^{43}~\rm{}ergs~s^{-1}$ is
$\simeq\!60$~\AA~\citep{xu_baldwin_08}. Other 
sightlines above the wind cone ($0^{\circ} \leq i \leq
70^{\circ}$) yield qualitatively similar spectra. Overall, the
presence of the right ``sort'' of emission lines is encouraging, but
their weakness is an obvious shortcoming of the model. We will discuss
the topic of line emission in more detail in Section \ref{line_weakness}.

\subsubsection{Sightlines looking through the base of the wind cone}
\label{base_cone}

Let us finally consider the highest inclinations, which correspond to
sightlines that do not lie along the wind cone, but for which the 
UV continuum source (i.e. the central region of the
accretion disk) is viewed through the dense, low-velocity base of the
outflow. It is not obvious what type of QSO one might expect to see
from such sightlines, since in the standard model of QSOs, one might 
expect the torus to obscure such sightlines. 

\begin{figure}
\plotone{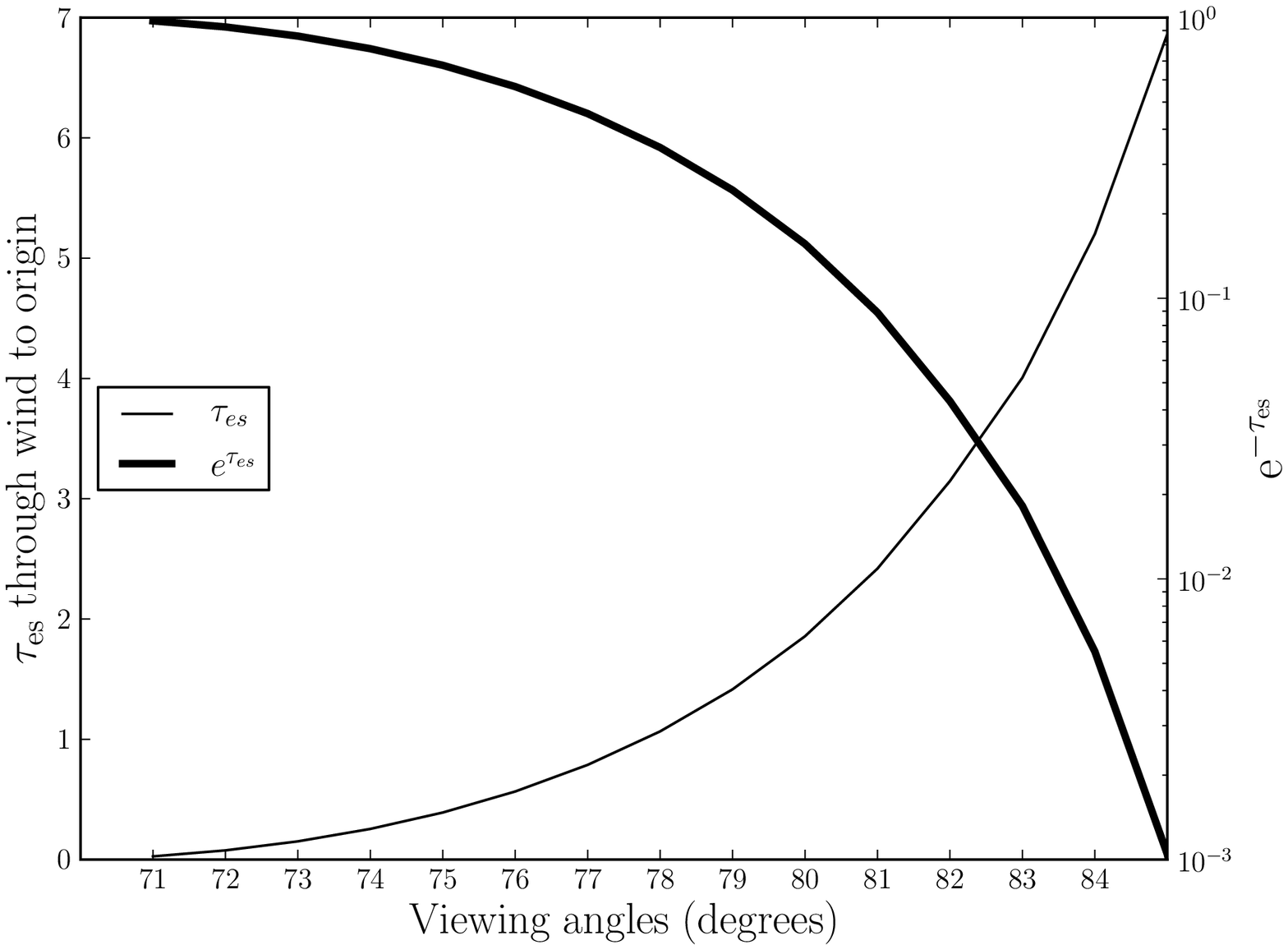}
\caption{Electron scattering optical depth through the wind towards the origin. The heavy line is the expected attenuation.}
\label{thompson_tau}
\end{figure}

Figure \ref{thompson_tau} shows that the $\tau_{es} > 7$ for this
inclination, and indeed we find that essentially all of the 
radiation emerging from the model in this direction has been scattered
several times within the wind before ultimately escaping along this
vector. This is in line with results presented by \cite{sim_2010}, who
found that, for Compton-thick winds, AGN spectra at high inclination
are dominated by scattered radiation. Recent observational
work \citep[e.g.][]{treister_09} has demonstrated that there is a population of 
AGN in the local universe where the X-ray source is completely obscured, 
and the X-ray spectrum is dominated by reflection. These high inclination sightlines
could represent these so called ``reflection dominated Compton thick AGN''.

The shape of the predicted emission features is dominated by the
rotational velocity field in the wind. Taking C~\textsc{iv} as an 
example and examining Figure \ref{physical_wind}, we see that the
fractional abundance  
peaks in a region where the rotational velocity is significantly
higher than the outflow velocity. Scattering from this region (along
with any thermal emission) will produce the double-peaked line profile 
that is characteristic of line formation in rotating media 
\citep[e.g.][]{smak_81,marsh_88}. The blue part of this profile is
then likely suppressed, since it is superposed on and/or absorbed by 
blue-shifted absorption in the base of the wind. 

\section{Discussion}
\label{discussion}

The results shown in the previous section confirm that a simple disk
wind model can produce the characteristic BAL features seen in about
$\simeq 20\%$ of QSOs. However, they also reveal some significant
shortcomings of the model. In this section, we will consider some of
the key questions raised by our results in more detail and point out
promising directions for future work.

\subsection{The Sensitivity of BAL Features to X-ray Luminosity}
\label{xray_sen}

\begin{figure}
\plotone{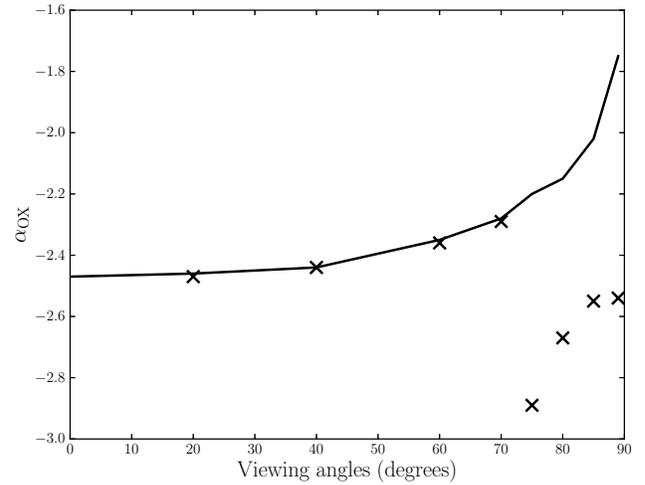}
\caption{Emergent $\alpha_{OX}$ values for benchmark model (symbols) and without wind (solid line)}
\label{alpha_ox_vs_theta}
\end{figure}

As noted in Section~\ref{xray_spec_lum}, our benchmark model
is rather X-ray weak compared to most real (BAL)QSOs. To
quantify this, Figure~\ref{alpha_ox_vs_theta} shows the
viewing angle dependence of $\alpha_{OX}$ for our benchmark
model. Clearly, $\alpha_{OX}$ (i.e. the X-ray to optical
ratio) is a strong function of inclination. Two effects are
responsible for this.

First, the UV and optical fluxes are dominated by the accretion disk
and thus decline steeply with inclination, due to foreshortening and
limb-darkening. By contrast, the X-ray source is assumed to emit
isotropically (although the disk can obscure the ``lower''
hemisphere). Thus, in the absence of any outflow, $\alpha_{OX}$
increases with inclination, as the X-rays become relatively
more important. This is the trend marked by the solid line in
Figure~\ref{alpha_ox_vs_theta}.

Second, sightlines to the central engine that cross the wind
cone can be affected by absorption and radiative transfer in the
outflow. In particular, bound-free absorption preferentially
suppresses the X-ray flux along such sightlines in our model, and
hence reduces $\alpha_{OX}$. The crosses in
Figure~\ref{alpha_ox_vs_theta} show the predicted values of
$\alpha_{OX}$ taking both effects into account. This shows that the
suppression of $\alpha_{OX}$ is strongest for sightlines looking
directly into the wind, i.e. the same viewing angles that produce
BALs. 

For our
benchmark model, we find that $\alpha_{OX} = -2.44$ for $i=40^\circ$
and $\alpha_{OX} = -2.89$ for $i=75^\circ$. If these viewing
angles can be taken to represent non-BAL QSOs and BALQSOs respectively, 
we can compare these values to those
expected from the observed scaling relation between 
$L_{2500\mbox{\scriptsize{\AA}}}$ and $\alpha_{OX}$ found by \cite{just_07} for
non-BAL QSOs,
\begin{equation}
\alpha_{OX}=(-0.140\pm0.007)\log(L_{2500\mbox{\scriptsize{\AA}}})+(2.705\pm0.212).
\end{equation}
Taking the values of $L_{2500\mbox{\scriptsize{\AA}}}$ for the two sightlines from our
benchmark model, we obtain empirical predictions of $\alpha_{OX} =
-1.66\pm0.43$ for $i=40^\circ$ and $\alpha_{OX} = -1.52\pm{0.43}$ for
$i=75^\circ$. However, we still have to correct for the fact that
BALQSOs are known to be X-ray weak (presumably due to absorption in
the outflow). \cite{gibson_09} find that the median $\Delta\alpha_{OX}
= \alpha_{OX,BALQSO} - \alpha_{OX,QSO} = -0.17$ for all BALQSOs with
$BI > 0$, but the distribution of $\Delta\alpha_{OX}$ is quite wide,
and systems with large $BI \gtappeq 1000~{\rm km~s}^{-1}$ tend to have
larger $\Delta\alpha_{OX}$. If we adopt $\Delta\alpha_{OX} \simeq -0.3$
as typical for strong BALQSOs (like our benchmark model), the empirically
predicted value for the $i=75^\circ$ BALQSO sightline becomes
$\alpha_{OX} \simeq -1.8\pm0.4$ (where the quoted uncertainty is
purely that associated with the scaling relation for non-BAL QSOs). 

These numbers make the X-ray weakness of the benchmark model clear: 
at fixed $L_{2500\mbox{\scriptsize{\AA}}}$, the emergent X-ray flux predicted
by the model is at least 2 orders of magnitude lower than the average
observed values. Equivalently, the model corresponds to a $2-3\sigma$
outlier in the $\alpha_{OX}$ distribution of (BAL)QSOs. 

\begin{figure*}
\plotone{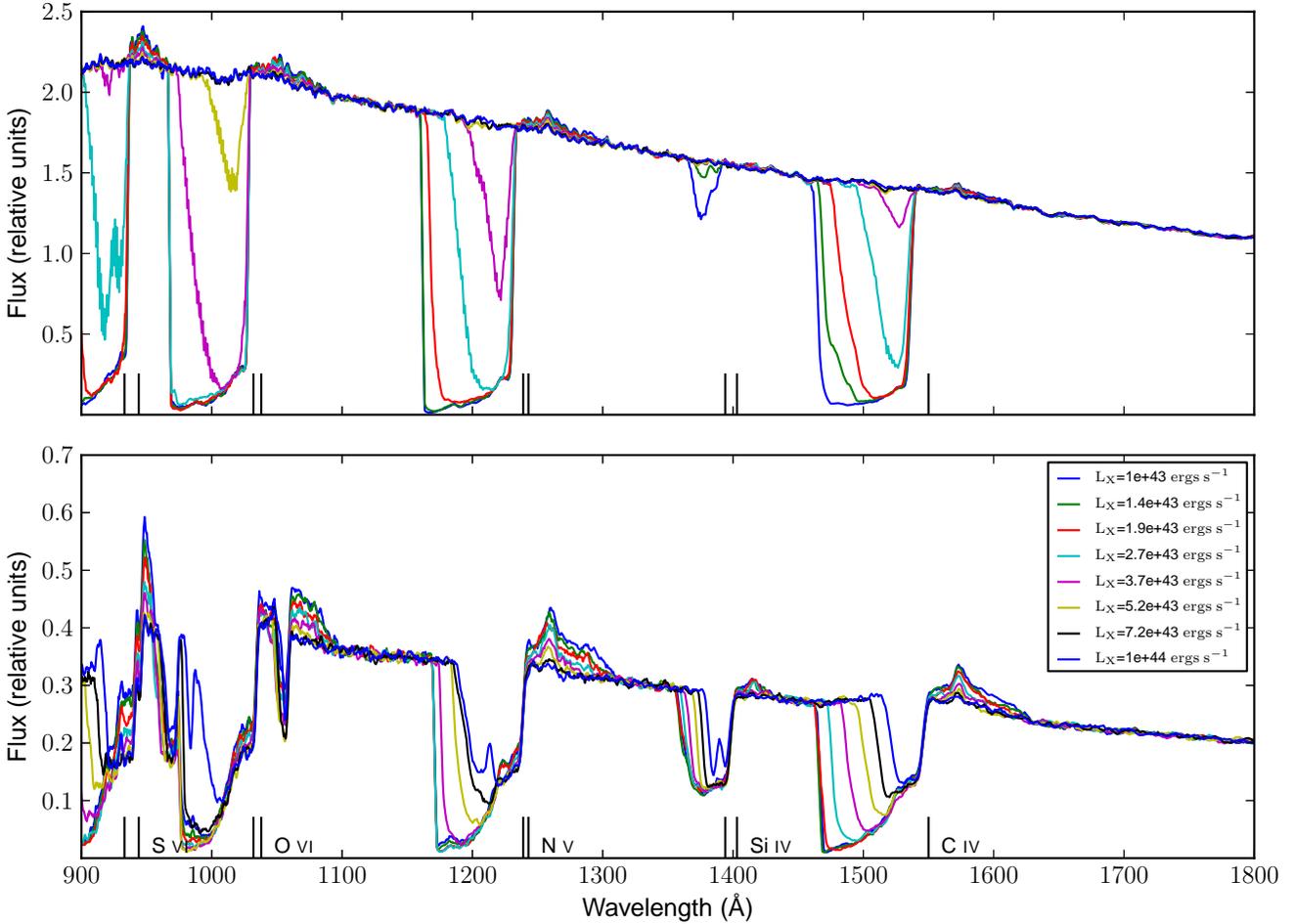}
\caption{Spectra at $75^{\circ}$ (top panel) and $80^{\circ}$ 
 sightlines for the benchmark model with a range of
  2-10~keV luminosities}
\label{xray_luminosity}
\end{figure*}

To investigate whether we can increase $L_X$ substantially while
retaining the observed BAL features for sightlines looking into the
wind cone,
Figure \ref{xray_luminosity} shows the behaviour of the
emergent spectra for $i=75^\circ$ and $i=80^\circ$ as the X-ray
luminosity is increased from $L_X = 10^{43}~{\rm erg~s^{-1}}$ to $L_X
= 10^{44}~{\rm erg~s^{-1}}$. We will focus on the behaviour of
C~\textsc{iv} and O~\textsc{vi} as 
representative examples. For $i = 75^{\circ}$, the C~\textsc{iv} BAL 
feature becomes narrower with increasing $L_X$: it gradually disappears,
starting from the blue
edge. The feature is lost entirely by 
$L_{X}=5\times10^{43}\rm{~ergs~s^{-1}}$. The effect on the 
O~\textsc{vi} feature is more abrupt, with almost no effect up to a
luminosity of $3.7\times10^{43}\rm{~ergs~s^{-1}}$, but complete
disappearance for twice that luminosity. Similarly, for $i =
80^{\circ}$, C~\textsc{iv} weakens gradually in models with larger $L_{X}$,
while in this case O~\textsc{vi} shows hardly any response to $L_X$ until we reach  $L_{X}= 10^{44}\rm{~ergs~s^{-1}}$. Note that
C~\textsc{iv} is still present with a clear BAL signature at this
X-ray luminosity for $i = 85^{\circ}$.

\begin{figure*}
\plotone{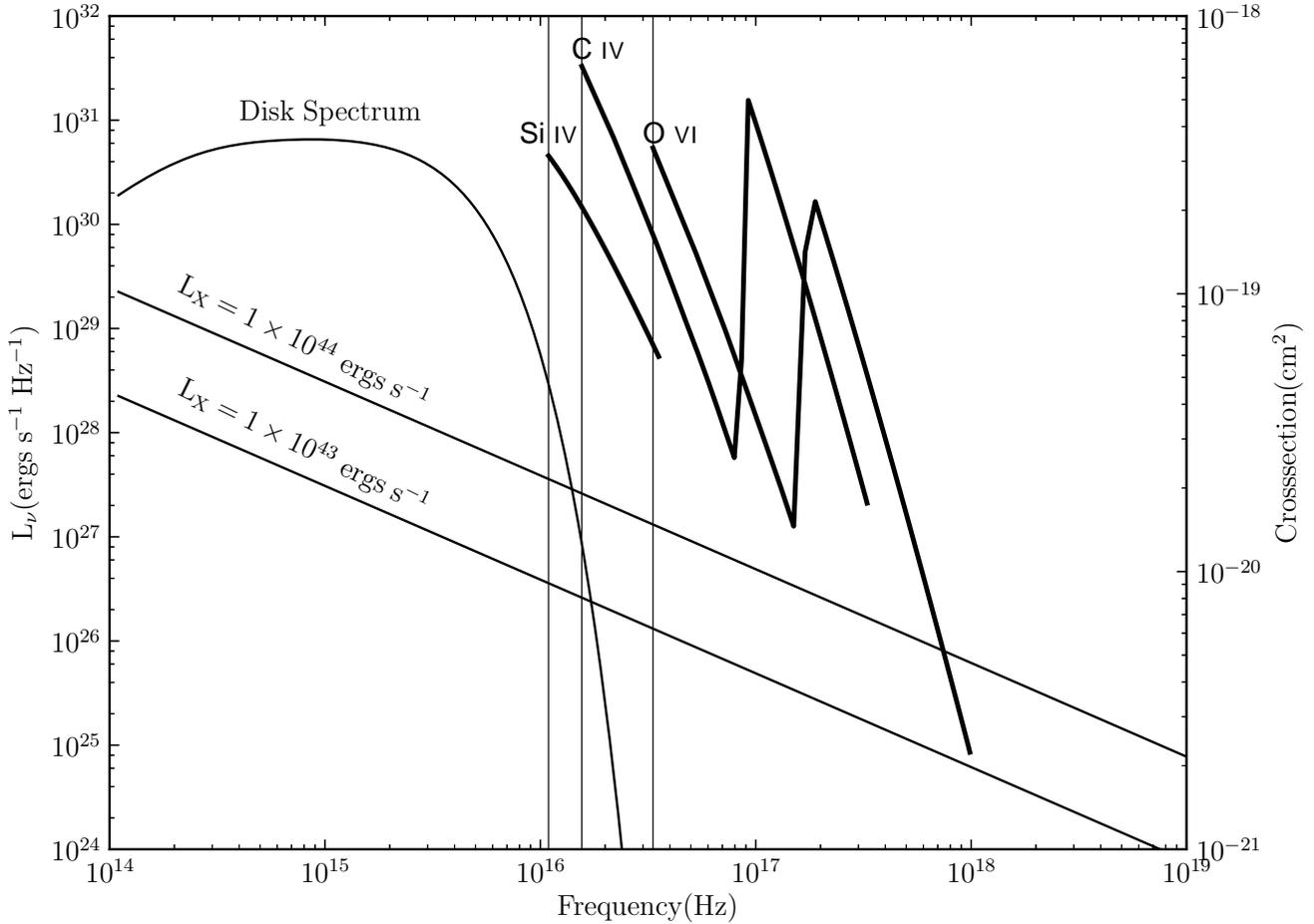}
\caption{Input spectrum from the disk component in the benchmark model,
  together with the contribution from the central source for two
  different values of $L_{X}$. The photoionization cross
  sections for C~\textsc{iv}, O~\textsc{vi} and Si~\textsc{iv} are
  also plotted over the same frequency range.}
\label{spec_csec}
\end{figure*}

The strong sensitivity of the UV resonance transition to $L_X$
in this range can be understood by considering the location of the
photoionization edges for these species. Figure \ref{spec_csec} shows
these locations relative to the angle-averaged 
disk spectrum and the X-ray spectra corresponding to $L_X = 
10^{43}~{\rm erg~s^{-1}}$ and $L_X = 10^{44}~{\rm erg~s^{-1}}$. 
The ionization edge of C~\textsc{iv} falls
very close to the frequency above which the X-ray source
takes over from the disk as the dominant spectral component. In fact,
for $L_X = 10^{43}~{\rm erg~s^{-1}}$, photo-ionization of
C~\textsc{iv} is driven primarily by disk photons, while for 
$L_X = 10^{44}~{\rm erg~s^{-1}}$, it is mainly due to photons from the 
X-ray source. The photo-ionization rate of higher-ionization species
like O~\textsc{vi} is dominated by the X-ray source even at $L_X =
10^{43}~{\rm erg~s^{-1}}$.

We conclude that moderate increases in $L_X$ (by a factor of a few)
would be feasible without any other parameter adjustments, but larger
changes will begin to destroy BAL features for more and more
sightlines. 
However, it is reasonable to anticipate that the negative
effects of increasing 
$L_X$ on BAL features may be mitigated by simultaneously increasing
the mass-loss rate in the wind, $\dot{M}_{wind}$. We have therefore carried
out a 2-dimensional sensitivity study, in which we vary both $L_X$ and
$\dot{M}_{wind}$. The results -- focussing particularly on the key
C~\textsc{iv} line profile for $i = 75^{\circ}$ -- are shown in
Figure \ref{CIV_vsmdot}. In each panel of this figure, we also give
the BI measured for the profile shown, as well as values of $F/F_0$;
the ratio of the 
  continuum at $\rm{-30000km~s^{-1}}$ to that
  obtained in a model without a wind.

The second column in Figure~\ref{CIV_vsmdot} illustrates the effect of 
increasing ${L_{X}}$ while keeping the mass-loss rate at the
benchmark value of $\dot{M}_{wind} = 5 {\rm M_{\odot}~yr^{-1}}$ (as in
Figure \ref{xray_luminosity}). We see again that the C~\textsc{iv}
BAL feature persists up to at least $L_X = 10^{44}~{\rm erg~s^{-1}}$,
although it becomes both weaker and narrower with increasing $L_X$. 
For higher mass-loss rates (third and beyond columns in
Figure~\ref{CIV_vsmdot}), the C~\textsc{iv} profile becomes
increasingly insensitive to $L_X$. So, as expected, increasing
$\dot{M}_{wind}$ allows for the production of strong BAL features in
species like C~\textsc{iv} even at higher irradiating $L_X$. 
Thus one way to address the X-ray weakness of the benchmark model
(i.e. increase $\alpha_{ox}$) is to increase both $L_X$ and
$\dot{M}_{wind}$. 

This remedy has a side effect: any increase in $\dot{M}_{wind}$ also
implies an increase in $\tau_{es}$, the electron-scattering optical
depth through the outflow. This reduces the {\em observed}
X-ray and and optical luminosity of the BALQSO. For moderate
inclinations and mass-loss rates, the emergent continuum flux will
scale simply as $e^{-\tau_{es}}$, although for the highest
inclinations and $\dot{M}_{wind}$, scattered radiation will dominate
the continuum and break this scaling. In any case, our main conclusion
is that, for high mass-loss rates, we are able to produce BALs over a
wider range of $L_{X}$ but at the expense of significant continuum
suppression. If this is to provide a realistic route towards
geometrical unification, it implies that the {\em intrinsic}
luminosities of BALQSOs must be higher than suggested by their 
observed brightness.

There are, of course, other possibilities. For example, if the outflow
is clumpy \citep[e.g.][]{krolik_81,arav_99}, some of 
the X-ray radiation may pass through the wind 
unimpeded, thus increasing the observed $L_X$ for sightlines inside
the wind cone. Such a solution would have the added advantage that the
higher density within the clumps would naturally result in a lower
ionization state for a given X-ray flux. Relaxing the assumption of a
smooth flow would, of course, take us back in the direction of cloud
models of the BLR and BALR. We plan to investigate the pros and cons of
such scenarios in future work.

\subsection{The Weakness of Line Emission Produced by the Outflow}
\label{line_weakness}

As mentioned in Section \ref{above_cone}, our benchmark model does not
produce strong emission line features, especially at low inclination
models where the geometric unification scenario suggests Type I QSOs
should be seen. This suggests that modifications are required in some
aspect of the structure or the physics that is incorporated into
\textsc{python}, if the geometric unification scenario is correct.

In considering this shortcoming, it is interesting to compare with the disk
wind modelling carried out by \citealt{murray_95} [hereafter
MCGV95]. They explicitly model a 
line-driven outflow, and the broad characteristics of their disk wind
are sufficiently similar to our benchmark model to make a comparison
interesting. MCGV find that their outflow 
produces BAL troughs when viewed at high inclinations. However, they
also find that their model produces sufficient collisionally excited
emission in species like C~\textsc{iv} to explain the BEL in both
BALQSOs and non-BAL QSOs. In their model, this C~\textsc{iv} emission
arises in a layer of the wind close to the disk plane where 
C~\textsc{iv} is the dominant ionization stage of carbon. They argue
that this region extends from $\simeq 10^{16}~{\rm cm}$ to $\simeq 10^{17}~{\rm cm}$
and is characterized by $T_{e} \simeq 20,000~{\rm K}$ and $n_e \simeq
10^{10}~{\rm cm^{-3}}$. It is the integrated, collisionally excited
emission from this region that dominates the C~\textsc{iv} line flux
in their model. 

In the case of our benchmark model, Figure~\ref{physical_wind} shows 
that the fractional C~\textsc{iv}
abundance is high ($f_{\rm C~\textsc{iv}} \gtappeq 0.1$) in a region
extending from $x \simeq 5 \times 10^{16}~{\rm cm}$ to $x \simeq 3
\times 10^{17}~{\rm cm}$. However, the density in this region is
considerably lower than in the C~\textsc{iv} line-forming region
considered by MCGV: $10^7~{\rm cm^{-3}} \ltappeq n_e \ltappeq
10^9~{\rm cm^{-3}}$. It is this lower density that almost certainly 
explains the difference in C~\textsc{iv} emissivity in the models.
In a later version of their model \citep{murray_chiang_98},
the production of C~\textsc{iv} is dominated by lower density ($n_e
\simeq 10^8~{\rm cm^{-3}}$) material lying at larger radii $r \gtappeq
10^{18}~{\rm cm}$. In these models, it is the large emitting volume
that explains the strength of collisionally excited C~\textsc{iv}
feature.

\begin{figure*}
\plotone{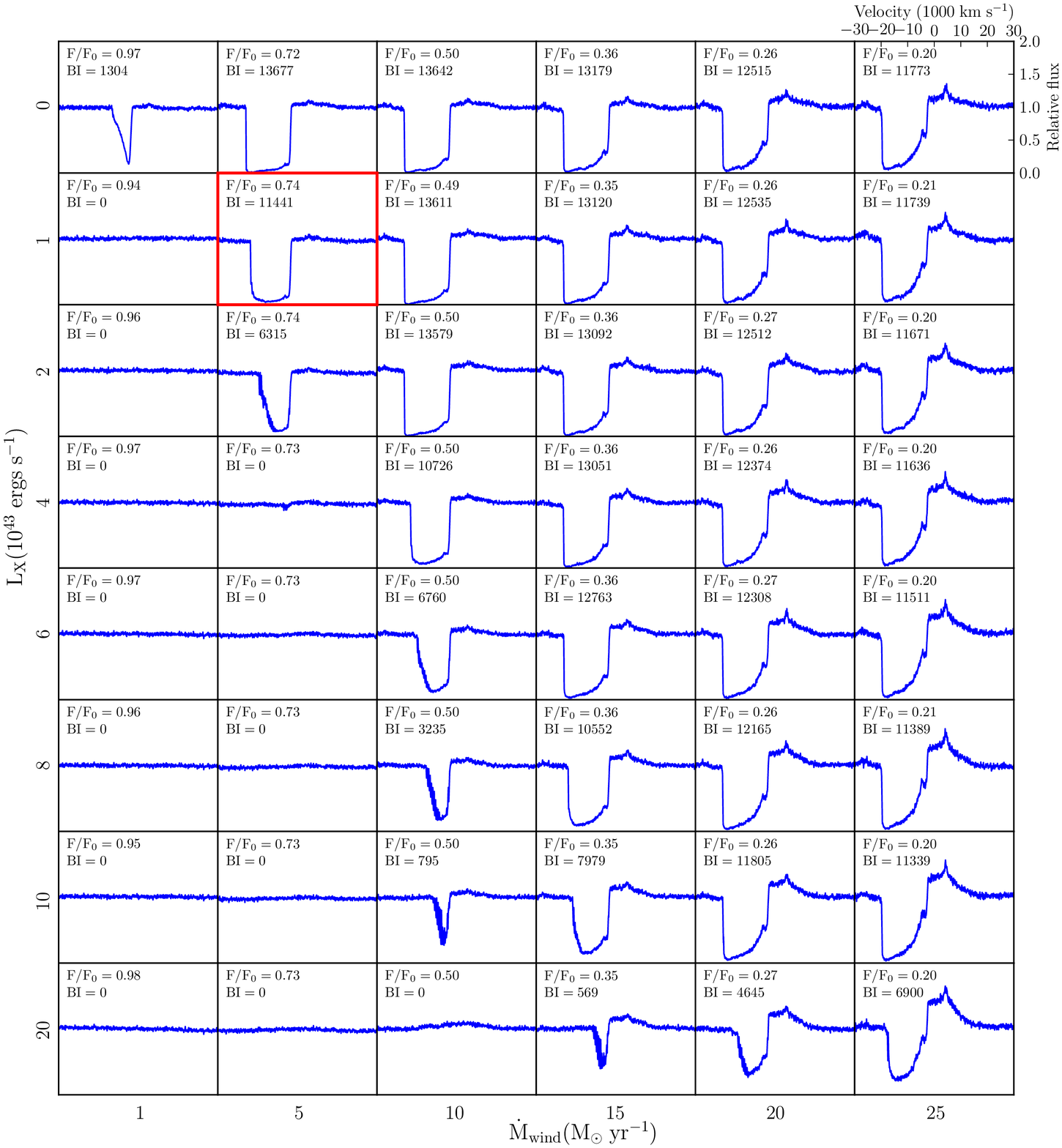}
\caption{Spectrum detail around the C~\textsc{iv} line feature for the
  benchmark model and others with a range of Xray luminosity and
  $\rm{\dot{M}_{wind}}$. Spectra are all for $i=75^{\circ}$ and are
  normalised to a linear continuum model fitted at
  $\rm{\pm30000km~s^{-1}}$. Values of $F/F_0$ show the ratio of the 
  continuum at $\rm{-30000km~s^{-1}}$ to that
  obtained in a model without a wind. The scales for each subplot are
  given on the upper right hand subplot and the baseline model is
  highlighted with a red border.}
\label{CIV_vsmdot}
\end{figure*}

These comparisons suggest that moderate changes to the benchmark 
model may be enough to produce BELs via collisionally excited line
emission in the outflow. It is likely that such changes would go in
the direction of increasing the density in some parts of the wind,
either by modifying existing wind parameters or by introducing
clumpiness. In fact, we can clearly see that the
emission component of the C~\textsc{iv} line in Figure~\ref{CIV_vsmdot} does
increase with increasing $\dot{M}_{wind}$. However, even our highest
$\dot{M}_{wind}$ models do not yet produce enough emission at low
inclination angles. Thus, while the benchmark model shows some
promise in the context of geometric unification, it remains to be seen
whether (and how) it can be modified to produce the strong emission
lines seen in both BALQSOs and non-BAL QSOs. This will be the subject
of a separate paper.

Before we leave this section, it is interesting to ask why the {\em
scattering} of C~\textsc{iv} photons in the outflow does not
produce emission lines of sufficient strength in our model. In CV
winds, for example, modelling suggests that resonance
scattering is the primary mechanism for producing the observed
emission line profiles (\citealt{slos_vit,kwd}, LK02, \citealt{noebauer_10}).
The critical factor in understanding the relative importance of
scattered photons to the line formation process is the combination of
foreshortening, limb-darkening, outflow orientation and covering factor.
The outflows in CVs are thought to be ejected roughly perpendicular
to the accretion disk, albeit over quite a large range of opening angles. 
Moreover, CV winds emerge directly above the continuum emitting
regions of the disk, 
whereas BALQSO winds cover only $\simeq 20\%$ as seen from the
continuum source. Thus, in a CV, an outflow that is optically thick in
C~\textsc{iv} over some velocity interval will intercept virtually all
of the continuum photons within this interval -- and scatter them
into other sightlines, where they emerge as apparent line
emission. Conversely, our QSO wind scatters only a small fraction of 
the continuum photons, even in frequency intervals in which the wind is
optically thick. As a result, scattered photons will not contribute as much
to emission line formation.

\subsection{Ionization Stratification and Reverberation Mapping}

In our benchmark model, the ionization state of the outflow tends to
increase with increasing distance along poloidal streamlines
(Figure~\ref{physical_wind}). This is easily understood. Far away from
the central engine, in the region of the 
flow where the BAL features are produced, the wind geometry
approximates an optically thin, spherical outflow. In such an outflow,
$U \propto v(r)$, so as long as the wind accelerates, the
ionization parameter must also increase outwards.

This trend may appear to contradict reverberation mapping studies of
the BLR, which suggest ionization 
stratification in the opposite sense. For example, the C~\textsc{iv}
BLR radius at fixed UV luminosity inferred by \cite{kaspi_07} is
$\simeq 2\times$ smaller than that obtained by \cite{kaspi_05} 
for H$\beta$. This sense of ionization stratification is also
supported by recent microlensing results \cite[e.g.][]{guerras_13}.

However, our model is not necessarily in conflict with these empirical
findings. As discussed in Section~\ref{line_weakness}, it is clear that
the parts of the wind producing BALs (where $U$ increases outwards)
are unlikely to be co-spatial with the parts of the wind producing
BELs. More specifically, only the dense base of the disk 
wind is likely to produce significant amounts of collisionally excited
line emission. Thus, in the context of disk wind models, the reverberation
mapping results are unlikely to probe the ``radial'' ionization
structure of the outflow, but may instead probe the stratification
along the base of the wind in the $x$ direction. Here, the ionization
parameter can (and does) decrease outwards in our disk wind models.

\subsection{Implications}

Here we consider implications of our
benchmark model for AGN feedback in galaxy evolution 
and identifying possible driving mechanisms for the wind.

In energetic terms, effective feedback -- sufficient to establish the
$M_{BH}-\sigma$ relation, for example -- seems to require $L_k/L_{bol}
\simeq 0.005 - 0.05$ \citep{dimatteo_05, hopkins_elvis}. Here,
$L_{k}=\frac{1}{2}\dot{M}_{wind}v^2_{\infty}$ is the kinetic luminosity
of the outflow. Our
benchmark model is characterized by $v_{\infty} \simeq 20,000~{\rm
km~s^{-1}}$ and $\dot{M}_{wind}=5~{\rm M_{\odot}~yr^{-1}}$, which yields
$L_k/L_{bol} \simeq 0.025$. Thus an outflow of this type would be capable
of providing significant amounts of energy-driven feedback. 
The kinetic luminosity of the benchmark model is also
broadly in line with the empirical scaling between $L_{bol}$ and
$L_{k}$ found by \cite{king_13} for black holes across the entire
mass-scale, from compact binary systems to AGN/QSO. 

\cite{king_03,king_05,king_10}  has argued that QSO outflows interact with their host
galaxy in a {\em momentum-}, rather than energy-driven
fashion. In this case, the strength of the feedback an outflow can
deliver depends on its momentum flux, $\dot{M}_{wind}v_{\infty}$. This
scenario naturally produces the observed $M_{BH}-\sigma$ relation if
the outflows responsible for feedback satisfy $\dot{M}_{wind}v_{\infty} 
\simeq L_{Edd}/c$; this is the single-scattering limit for momentum
transfer in a radiatively-driven outflow from a QSO radiating at
$L_{edd}$. For our benchmark model, we find $\dot{M}_{wind}v_{\infty}
\simeq 0.7 L_{bol}/c \simeq 0.1 L_{Edd}/c$. In other words, the
momentum flux in our model is close to the single-scattering limit for 
its bolometric luminosity. Since black hole growth and cosmological
feedback are thought to be dominated by phases in which $L_{bol}
\simeq L_{Edd}$ 
\citep{soltan_82,yu_tremaine_02,dimatteo_05},
outflows of this type would probably also meet momentum-based
feedback requirements. 

The momentum flux in the outflow is also a key parameter for any disk
wind driven by radiation pressure, such as the line-driven winds
considered by MCGV95 and \cite{proga_stone_kallman}. In
particular, such winds typically satisfy the single-scattering limit
for momentum transfer from the radiation field to the outflow,
$\dot{M}_{wind} v_{\infty} \leq L_{bol}/c$. As noted above, our benchmark
model also satisfies this limit. However, this is somewhat
misleading: our disk wind subtends only $\simeq 20\%$ of the sky as seen
from the central source, and it intercepts an even smaller fraction
of the QSOs bolometric luminosity (due to foreshortening and limb-darkening 
of the disk radiation field).
The momentum flux in our
benchmark model actually exceeds the single-scattering
limit
if only momentum carried by photons that actually intercept the wind is
taken into account. Thus, in the context of radiatively-driven winds, either the 
mass-loss rate in our model must be overestimated or multiple
scattering effects must be important. The latter has been suggested
for quasi-spherical
Wolf-Rayet star winds \citep{lucy_abbott_93,springmann_94,gayley_95}
 but is much more challenging for the biconical wind considered here. 
Alternatively, the winds of QSOs may not be driven
(exclusively) by radiation pressure, with magnetic/centrifugal forces
providing the most obvious alternative \citep[e.g.][]{blandford_payne_82,
emmering_92,proga_03}.

Our discussion of all these considerations is preliminary: 
the benchmark model is far from perfect, and we have
not yet carried out a comprehensive survey of the available parameter
space. However, the first column in Figure~\ref{CIV_vsmdot} shows
that a reduction in the mass-loss rate from the benchmark value of 
$\dot{M}_{wind} = 5 {\rm M_{\odot}~yr^{-1}}$  to
$\dot{M}_{wind} = 1 {\rm M_{\odot}~yr^{-1}}$ virtually destroys the  
BAL features, even for $L_X = 10^{43}~{\rm erg~s^{-1}}$. Thus, at
least for our adopted wind geometry and kinematics, it seems that any
outflow capable of producing BAL features is also likely to provide
significant feedback and pose a challenge to line-driving.

\subsection{Outlook}

In our 
view, the most significant shortcomings of the benchmark model are its
intrinsic X-ray weakness and its inability to produce collisionally
excited BELs. Both problems might be resolved by moderate changes in
wind parameters, so an important step will be to explore the parameter
space of our model. We plan to adopt two complementary approaches in
this: (i) systematic grid searches; (ii) targeted
explorations guided by physical insight. The latter approach is
important, since compromises will have to be made in the 
former: a detailed and exhaustive search over all relevant parameters 
is likely to be prohibitive in terms of computational resource requirements. As an
example of how physical insight can guide us, we note that both the
X-ray and emission line problems might be resolved by increasing the
density in specific wind regions.

\section{Summary}
\label{conclusion}

This paper represents the first step in a long-term project to shed
light on the nature of accretion disk winds in QSOs. These outflows
may be key to the geometric unification of AGN/QSO and might also
provide the feedback required by successful galaxy evolution
scenarios. 

In this pilot study, we have focussed on the most obvious
signature of these outflows, the broad, blue-shifted absorption lines
seen in $\simeq 20\%$ of QSOs. We have constructed a kinematic disk
wind model to test if it can reproduce these features. This
benchmark model describes a rotating, equatorial and biconical
accretion disk wind with a mass-loss rate $\dot{M}_{wind} \simeq 
\dot{M}_{acc}$.\footnote{
Strictly speaking, this is not entirely self-consistent, since the presence of
such an outflow would alter the disk's temperature 
structure. However, we neglect this
complication, since the vast majority of the disk luminosity arises
from well within our assumed wind-launching radius.}
A Monte Carlo ionization and radiative transfer code,
\textsc{python}, was used to calculate the ionization state of the
outflow and predict the spectra emerging from it for a variety of
viewing angles. Our main results are as follows:

\begin{list}{}{}

\item[(i)] Our benchmark model succeeds in producing BAL-like features
for sightlines towards the central source that lie fully within the
wind cone. 

\item[(ii)] Self-consistent treatment of ionization and radiative
transfer is necessary to reliably predict the ionization state of the
wind and the conditions under which key species like C~\textsc{iv} can
efficiently form. 

\item[(iii)] The benchmark model does not produce sufficient
collisionally excited line emission to explain the BELs in
QSOs. However, we argue that moderate modifications to its parameters 
might be sufficient to remedy this shortcoming.

\item[(iv)] The ionization structure of the model, and its ability to
produce BALs, are quite sensitive to the X-ray luminosity of the
central source. If this is too high, the wind becomes overionized. In
our benchmark model, $L_X$ is arguably lower than indicated by
observations. Higher values of $L_X$ may require higher outflow
columns -- e.g. via higher $\dot{M}_{wind}$ -- in order to still produce
BALs.

\item[(v)] For our adopted geometry and kinematics, $\dot{M}_{wind}
\gtappeq \dot{M}_{acc}$ is required in order to produce significant
BAL features.  The kinetic luminosity and momentum carried by such
outflows is sufficient to provide significant feedback.

\end{list}

\section*{Acknowledgements} The work of NH, JHM and CK are supported by the Science
and Technology Facilities Council (STFC), via studentships and a
consolidated grant, respectively. The authors would like to thank the 
anonymous referee for useful comments.
\bibliographystyle{mn2e}

 \label{lastpage}

\end{document}